\documentclass[aps,prd,amsmath,preprint,superscriptaddress,nofootinbib]{revtex4-1}
\usepackage{geometry}
\usepackage[utf8]{inputenc}
\usepackage[T1]{fontenc}
\usepackage{verbatim}
\usepackage{amsmath,amssymb,amsfonts}	
\usepackage{slashed}
\usepackage{graphicx}
\usepackage{color}
\usepackage[dvipsnames]{xcolor}
\usepackage{hyperref} 

\usepackage[normalem]{ulem}

\usepackage{graphicx}
\usepackage{dcolumn}
\usepackage{bm}

\begin{document}

\preprint{FERMILAB-PUB-23-339-T}

\baselineskip 0.7cm

\bigskip

\title{Leveraging intermediate resonances to probe CP violation at colliders}

\author{Innes Bigaran}
\email[]{ibigaran@fnal.gov}
\affiliation{Physics \& Astronomy Department, Northwestern University, Evanston, IL}
\affiliation{Theoretical Physics Department, Fermilab, P.O. Box 500, Batavia, IL 60510, USA}

\author{Joshua Isaacson}
\email[]{isaacson@fnal.gov}
\affiliation{Theoretical Physics Department, Fermilab, P.O. Box 500, Batavia, IL 60510, USA}

\author{Taegyun Kim}
\email[ ]{tkim12@nd.edu}
\affiliation{Uichang District Office, 468, Dogye-dong
Uichang-gu, Changwon, 51381, South Korea}

\author{Karla Tame-Narvaez}
\email[]{karla@fnal.gov}
\affiliation{Theoretical Physics Department, Fermilab, P.O. Box 500, Batavia, IL 60510, USA}
\affiliation{{Physics Department, University of Notre Dame, 225 Nieuwland Science Hall, Notre Dame, South Bend, IN 46556, USA}}

\date{\today}

\begin{abstract}
We explore the phenomenological impact of interference in tree-level contributions to three-body final states in $2\to 3$ scattering processes. This work introduces a novel search strategy leveraging asymmetries to enable sensitivity to CP-violating effects in less well-explored regions of phase space. Analytically, we demonstrate the effectiveness of this observable in probing interference between Standard Model charged-current decays and effective left-handed vector interactions, illustrated in a toy model featuring a scalar leptoquark, $S_1 \sim (3, 1, -1/3)$. Numerically, we apply this framework to studying the process $pp\to b \tau\nu$; unlike traditional high-$p_T$ searches or ``bump hunts", this approach utilizes an intermediate energy regime -- where new physics is neither light enough to be produced on shell or heavy enough to justify an effective field theory treatment. A proof-of-principle analysis at parton level demonstrates a percent-level asymmetry, with sensitivity also to BSM weak-CP phase. While the specific phase sensitivity is diminished at particle level due to showering and detector effects, a machine learning classifier can recover sensitively to the presence of SM-BSM interference, significantly outperforming standard analysis methods. Notably discrimination between BSM signal and SM background could be achieved at the 2$\sigma$ level for the current LHC dataset and 8$\sigma$ at the High-Luminosity LHC. Moreover, this asymmetry observable as defined can also be more broadly applied to other searches for CP-violation in $2\to 3$ processes in present and future collider environments.
\end{abstract}

\maketitle

\section{Introduction}
\label{sec:intro}

Within and beyond the Standard Model~(SM) of particle physics, studying the nature of CP-violation~(CPV) provides a window to understanding  our universe. The SM flavor sector contains a single CP-violating parameter in the Cabibbo-Kobiyashi-Maskawa (CKM) quark-mixing matrix~\cite{Kobayashi:1973fv}.\footnote{The SM extended to include massive neutrinos introduces at least one additional CP-violating phase.} To explain the baryon asymmetry of the universe, the rate of CPV in the SM alone is insufficient. Furthermore, since it is not a symmetry of the SM there is no reason to think it would be exactly conserved in a more extended theory. Nonetheless, strong constraints from flavour and precision physics measurements constrain the extent of CPV in BSM models, motivating a conservative introduction of new sources: perhaps by introducing extended flavour symmetries or decoupling CPV new physics from the flavour sector. The latter in particular motivates probing the nature of CPV not only in low-energy probes such as meson-mixing and electric dipole moment searches, but also seeking direct signals of this asymmetry which may manifest in high-energy experiments. 

One may broadly separate the notion of irreducible CP phases into two kinds: ``weak'' phases and ``strong'' phases. ``Weak'' phases are CP\emph{-odd} and arise from explicitly complex Lagrangian parameters, e.g. the aforementioned CKM phase. ``Strong'' phases are CP\emph{-even}, typically arising from (QCD) dynamics, e.g. from the on-shell propagation of intermediate-state particles. To probe CP asymmetry in heavy particle decays one typically constructs the ratio of the sum and the difference of two decays related to one another by CP-conjugation, $\mathcal{A}_\text{CP}$. In general for $ \mathcal{A}_\text{CP}\neq 0$, the amplitude for the decay rate must contain at least two interfering amplitudes with different CP-even and CP-odd phases (see App.~\ref{App:interf}).\footnote{It is possible to avoid the condition of requiring amplitudes with different
CP-even phases by studying triple-product asymmetries \emph{if} the momenta and helicities of final-state particles can be determined, see e.g. \cite{Nowakowski:1988dx, Valencia:1988it, Korner:1990yx, Kayser:1989vw, Bensalem:2000hq,Datta:2003mj,Aguilar-Saavedra:2004tnj,Bartl:2004jj,Langacker:2007ur}.}

In Ref.~\cite{Berger:2011wh, Berger:2012gh}, the authors propose a new CPV observable in three-body decays, where the strong-phases contributing to $\mathcal{A}_\text{CP}$ are sourced by different virtualities of propagating tree-level mediators. Consider a process $\Phi_1(p_1) \to \Phi_a(p_a) \Phi_b(p_b) \Phi_c(p_c)$, with a decay rate $\Gamma$, this asymmetry is given by
\begin{align}
\label{eq:1to3}
\mathcal{A}_\text{CP}^{1\to 3} = \frac{\frac{d\Gamma}{dp_{bc}^2 dp_{ac}^2}-\frac{d\bar{\Gamma}}{dp_{bc}^2 dp_{ac}^2}}{\frac{d\Gamma}{dp_{bc}^2 dp_{ac}^2}+\frac{d\bar{\Gamma}}{dp_{bc}^2 dp_{ac}^2}}
\end{align}
where $\bar{\Gamma}$ is the decay rate of the CP-conjugate process, and $p_{ij}^2$ is the invariant-mass squared of the $\Phi_i- \Phi_j$ system in the final state. If there are distinct topologies contributing to this decay, where one contains a mediator with mass $M_1$ coupling to $\Phi_b- \Phi_c$, and the other with a mediator of mass $M_2$ coupling to $\Phi_a- \Phi_c$, then this asymmetry will vary over the $p_{bc}^2$- $p_{ac}^2$ plane as the propagators access different virtualities. The variation in this asymmetry may be studied utilizing the Dalitz distribution of the process and its CP-conjugate. 

In this paper, we note that this asymmetry definition may be extended to $2\to 3$ scattering processes to define $\mathcal{A}_\text{CP}^{2\to 3} $ -- replacing $\Gamma$ with a cross-section, $\sigma$, in Eq.~\eqref{eq:1to3} -- such that for a process $\Phi_1(p_1) \Phi_2(p_2)\to \Phi_a(p_a) \Phi_b(p_b) \Phi_c(p_c)$, 
\begin{align}
\label{eq:2to3}
\mathcal{A}_\text{CP}^{2\to 3} = \frac{\frac{d^2\sigma}{dp_{bc}^2 dp_{ac}^2}-\frac{d^2\bar{\sigma}}{dp_{bc}^2 dp_{ac}^2}}{\frac{d^2\sigma}{dp_{bc}^2 dp_{ac}^2}+\frac{d^2\bar{\sigma}}{dp_{bc}^2 dp_{ac}^2}}
\end{align}
which is a measure of CP-violation accessible in collider events. 
We stress that this asymmetry can and should be assessed for viability in various collider and model contexts, but for the purpose of this work we will focus on studying $\mathcal{A}_\text{CP}^{2\to 3} $ for the case of interference between a SM and BSM contribution to a fixed process. 
Note that to define such a CP asymmetry, the initial state must be a CP eigenstate.

As this asymmetry is sensitive to the relative virtuality between the two propagators, even in a kinematic region where one of these propagators is heavy enough to be naively parameterized in an EFT, this interference measure sampled around the alternate propagator can remain sensitive to new CP-odd phases. In the EFT, we have an effective Lagrangian 
\begin{equation}
    \mathcal{L}= \mathcal{L}_\text{SM} + \sum_{d=5}^N \sum_i\frac{c_i^{(d)}}{\Lambda^{d-4}} \mathcal{O}_i^{(d)}
\end{equation}
where $d$ represents the mass-dimension of the effective interaction from which any heavy BSM fields are integrated-out, and $\Lambda$ is the cutoff energy for the full theory. The coefficients $\{c_i^{(d)}\}$ represent the dimensionless Wilson coefficients~(WCs), parameterizing the effective interaction strengths at a fixed mass-dimension and energy scale.  At $d=6$, BSM contributions scale as $c/\Lambda^2$ and the squared-amplitude is given by
\begin{align}
|\mathcal{M}|^2=|\mathcal{M}_\text{BSM}|^2 + 2 \text{Re}( \mathcal{M}_\text{SM} \mathcal{M}_\text{BSM}^*)+|\mathcal{M}_\text{SM}|^2 \approx \frac{c^2}{\Lambda^4} + 2 \text{Re}( \mathcal{M}_\text{SM} \frac{c^*}{\Lambda^2})+|\mathcal{M}_\text{SM}|^2.\ \label{eq:BSMSM}
\end{align}
In the EFT framework, this interference effect is only nonzero if the BSM effect enters with the same Lorentz structure as  already present in the SM.
If $\Lambda$ is large, the interference term is less suppressed than the pure-BSM term and an observable isolating this term may provide a more sensitive probe for the effects of high-scale new physics. Furthermore, the term $\text{Re}( \mathcal{M}_\text{SM} \frac{c^*}{\Lambda^2})$ is sensitive to the phase-difference between the SM and BSM contribution (see App.~\ref{App:interf} for more detailed discussion), and therefore a CP-asymmetry isolating this effect is sensitive to phase information about the high-energy theory even in an EFT study.

Typically when searching for evidence for BSM at high-energy experiments, one proceeds in either of the following two ways: performing direct searches for on-shell resonances, or by integrating-out heavy new physics and constraining higher-dimensional effective operators. These two search methods are complementary and may be employed together as tools to constrain BSM effects. Above we have discussed interference in the EFT, when the dynamics of the BSM contact-interaction cannot be resolved. However, for $\mathcal{A}_\text{CP}^{2\to 3} $ the variation in virtualities of intermediate states mean that the full kinematic spectrum of final-state momenta can be utilized to test a hypothesis of BSM physics.  In particular, it may be sensitive to a kinematic region where an intermediate BSM state is not heavy enough to be integrated out, but not light enough to be produced on-shell
-- an intermediate region of energies, overlooked in collider searches for EFT effects or new resonances. 

There is a long history of studying interference between CP-phases in SM and BSM contributions to collider signatures. For example, in searches for $H \to c\bar{c}$ via exclusive Higgs decays~\cite{PhysRevD.37.1801, PhysRevLett.90.252001}, in high-energy diboson searches exploiting the interference between the SM and BSM contributions from $d=6$ EFT~\cite{Farina:2016rws,Bishara:2020pfx}, searches for top-quark flavour-changing neutral currents~\cite{Cremer:2023gne}, and utilization of the $J/\Psi$ resonance to search for BSM~\cite{Becirevic:2020ssj}. The asymmetry $\mathcal{A}_\text{CP}^{1\to 3}$ has been studied as a phenomenological probe of exotic Higgs decays~\cite{Chen:2014ona} and color octet scalars~\cite{He:2011ws}. Study of CPV in tree-level scattering with heavy fermion propagators was performed in Refs.~\cite{Pilaftsis:1989zt,Pilaftsis:1997dr}. Complementary to this previous work in the literature, here we propose this asymmetry $\mathcal{A}_\text{CP}^{2\to 3}$ as a probe of SM-BSM interference in collider scattering events. 

This manuscript is organized as follows. In Sec.~\ref{sec:int}, we present the significance of the interference between SM and BSM contributions in the era of the HL-LHC. Specifically, we focus on the $pp \to d_i \ell \nu_\ell$ process, discussing it in terms of two SM extensions: an EFT and the inclusion of an isospin-singlet scalar leptoquark (LQ), $S_1$. We also highlight the importance of the W and leptoquark mass poles in each scenario. In Sec.~\ref{sec:cpv_bsm}, we explore CP-violation in the $b \to c\tau\nu$ process, defining the weak and strong CP phases within the leptoquark model. Sec.~\ref{sec:analysis} contains an analysis for $pp$ colliders and presents our results. Finally, we conclude in Sec.~\ref{sec:clonclusion}.

\section{\texorpdfstring{SM/BSM interference in $2\to 3$ scattering}{Interference between SM and BSM in high-energy 23 scattering}}
\label{sec:int}

To begin, we first highlight a benefit of probing BSM-SM interference at a collider in the face of increased luminosity. The high-luminosity LHC upgrade (HL-LHC) is expected to provide approximately $3000$ fb$^{-1}$ of data at $\sqrt{s}=14$ TeV, an upgrade from the present searches at LHC Run-2 which were run at $\sqrt{s}=13$ TeV with an integrated luminosity of $139$ fb${}^{-1}$. Improved sensitivities are expected with increased luminosity due to a reduction in statistical uncertainties, 
and an observable which scales with $c$  rather than $c^2$ in Eq.~\eqref{eq:BSMSM} would yield more favorable improvement~\cite{Cremer:2023gne}. If the cross-section for the signal is dominated by the BSM-only term, then the sensitivity for the value of this parameter scales $\propto c^2/\sqrt{L}$ 
and thus the 95\% confidence interval on the value of $c$ scales as $1/\mathcal{L}^{1/4}$.
Therefore, the significance improvement from the present run to the end of the HL-LHC  is roughly a factor of $({3000/139})^{1/4}\approx 2.16$.
On the other hand, if an observable is dominated by BSM-SM interference, then the sensitivity to the WC scales like $c$, and the 95\% confidence interval of $c$ is proportional to $1/\mathcal{L}^{1/2}$, which leads to a significance improvement of a factor of 4.65.

To provide a concrete example of the application of $\mathcal{A}_\text{CP}^{2\to 3}$, we will refine our study here to charged-current interactions. In the SM, these processes occur via a left-handed vector current mediated the $W$. In BSM models, the effective interaction is parameterized by
\begin{align}
    \mathcal{L}_\text{WET} \supset& -\frac{4G_F}{\sqrt{2}}V_{ij}(C_{V_L})_{ji\ell}(\overline{u_i}\gamma^\mu P_L d_j)(\overline{\ell}\gamma_\mu P_L\nu_\ell)
\end{align}
where the EFT for the Weak Effective theory~(WET) is defined more completely in App.~\ref{appA:EFT}. Therefore, $\mathcal{A}_\text{CP}^{2\to 3}$ isolates the effect of a single Lorentz structure generated by BSM physics. This is in contrast to low-energy probes of CPV where they are often functions of different CP-violating effective interactions. We will apply $\mathcal{A}_\text{CP}^{2\to 3}$ to   $u_j g\to d_i \ell \nu_\ell$, as a contribution to the charge-asymmetry in the $pp\to d_i \ell \nu_\ell$ scattering process. This is largely inspired by the community interest in BSM models modifying the $b\to c \tau \nu$ interaction, to which this process provides a complementary probe~\cite{Altmannshofer:2017poe}.\footnote{ We study this process with the up-type quark in the initial state, although the alternative with a down-type quark in the initial state is also an interesting avenue of investigation but is beyond the scope of this work.} 

In practice, when working with high-energy interactions, one should re-derive the effective interactions in the SMEFT after running the WCs to the high-energy scale, but in practice here we will note that (a) $C_{V_L}$ has no anomalous QCD dimension, so the negligible running between the high- and low-energy scales for this coefficient can be neglected for the remainder of this work, (b) thus, the mapping between the SMEFT WC and the WET WC is a linear relation which is given in App.~\ref{appA:EFT} which corresponds to an effective relabeling of coefficients, which does not impact the physics results. For this reason, we retain WET WC labeling throughout this work. We draw the readers attention to App.~\ref{App:interf} for a more general treatment of the interference effects discussed in this section.

\begin{figure}[t]
\centering
\includegraphics[width=0.25\linewidth]{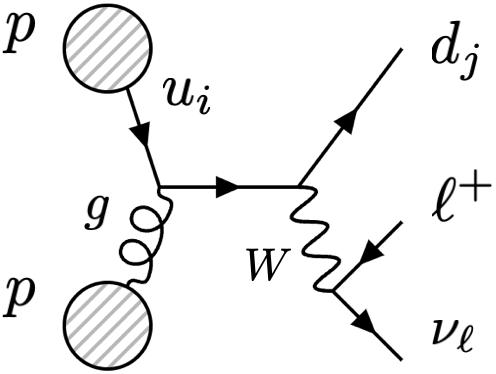} \hspace{1cm}
\includegraphics[width=0.25\linewidth]{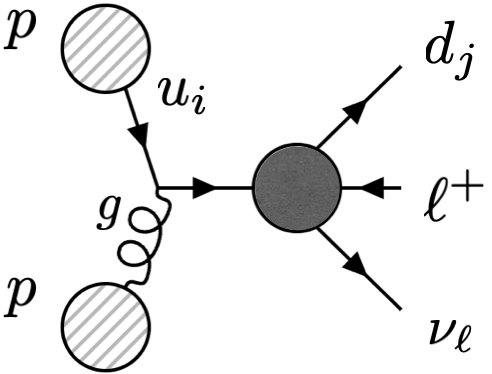} 
\\ 
\vspace{1cm}
\includegraphics[width=0.25\linewidth]{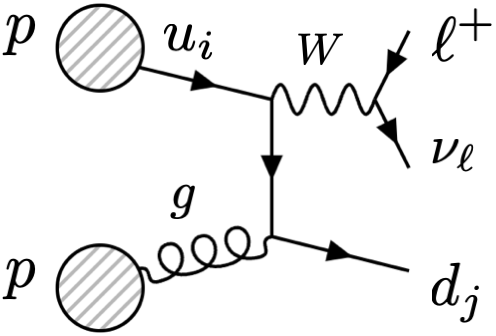}
\hspace{1cm}
\includegraphics[width=0.25\linewidth]{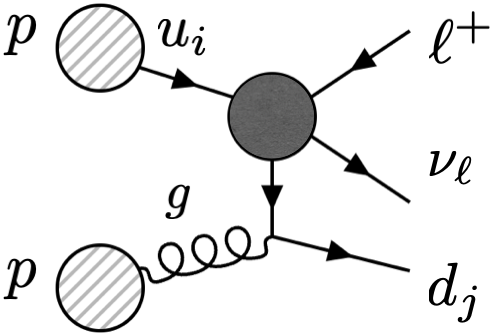}
\caption{SM (left) and BSM (right) contributions in the EFT to the $p(u_i)p(g)\to d_j \ell \nu_\ell$ process. Hashed circles refer to protons and the solid circles refer to the insertion of the dimension-six effective interaction, parameterized by the WC $C_{V_L}$. The dashed circles represent the proton initial states.}
\label{fig:pptaunu}
\end{figure}

\subsection{\texorpdfstring{Interference in $u_i g\to d_i \ell \nu_\ell$
}{Interference in ppdnul}}

We first consider the different topologies of the interactions in Fig.~\ref{fig:pptaunu} from the EFT perspective. The left diagrams show the SM contributions to the process $pp[u_i g] \to d_j \ell \nu_\ell$ and the right diagrams show BSM contributions to $pp[u_i g] \to d_j \ell \nu_\ell$ that arise from modifying the $u_i\to  d_j \ell \nu_\ell$ effective vertex (shown here as a solid circle). The top diagrams correspond to the initial-state gluon  interacting with the up-type quark, and the bottom diagrams to the initial-state gluon  interacting with the the down-type quark. Note that in this EFT treatment when limiting effective interactions to dimension six, we do not consider the case where a mediator is colored. This would permit an additional diagram where the gluon connects to the effective vertex. In a concrete model and in our numerical study, the effects of such a contribution will be considered, but for now we take this assumption to parameterize the dominant effect. 

In this work we will focus on three different levels of the calculation.
These are the \emph{partonic cross-section}, \emph{parton-level events}, and \emph{particle level events}.
The differential cross section is obtained through a convolution of the partonic cross-section with the parton distribution functions (PDFs), i.e.
\begin{equation}
    \frac{{\rm d}\sigma}{{\rm d}\Omega} = \int {\rm d}x_1 {\rm d}x_2 {\rm d}\Omega \sum_{i,j} f_{i,h_1}(x_1,\mu_F) f_{j,h_2}(x_2,\mu_F) \frac{{\rm d}\hat{\sigma}}{{\rm d}\Omega}.
    \label{eq:xsec}
\end{equation}
In the above equation, $f_{j,h}(x,\mu_F)$ is the PDF for obtaining a parton $j$ with momentum fraction $x$ from a hadron $h$ at a factorization scale of $\mu_F$, and $\frac{{\rm d}\hat{\sigma}}{{\rm d}\Omega}$ is the partonic cross section obtained from perturbative quantum field theory calculations. The factorization scale is an arbitrary scale that separates the definition of the PDF from the hard scattering process. When referring to parton-level events, we will use Monte-Carlo methods to sample momentum for the incoming and outgoing particles according to Eq.~\eqref{eq:xsec}. This will result in a set of ``events'' that contain contributions from the PDFs, but do not include effects of parton showers or hadronization. Finally, particle-level events are the obtained in the same way as parton-level events, but also undergo parton showers and hadronization. For a detailed summary about event generation see Ref.~\cite{Hoche:2014rga}.

To begin, we revisit the point that in order to define a CP asymmetry for this process (or a subprocess thereof) then the initial state \emph{must} be in CP eigenstate. This limits the choice of parton in the initial state solely to those with a sea quark, where the density of quark and anti-quark in the proton is identical. When considering this process beyond quark level, then the CP-asymmetry theoretically prescribed above cannot be extracted directly, and will provide a correction to the experimental \emph{charge} asymmetry, indicating the presence of CP-violating new physics. Details of this procedure will be discussed further in  Section~\ref{sec:analysis}.

In the $c g\to d_i \ell \nu_\ell$ subprocess, we define the partonic CP asymmetry as
\begin{align}
  \mathcal{A}_\text{CP}\equiv  \frac{{d\hat{\sigma}}/{d {q_{\ell \nu}^2}d {q_{d_j \nu}^2}}- {d\bar{\hat{\sigma}}}/{d {q_{\ell \nu}^2}d {q_{d_j \nu}^2}}  }{{d\hat{\sigma}}/{d {q_{\ell \nu}^2}d {q_{d_j \nu}^2}}+{d\bar{\hat{\sigma}}}/{d {q_{\ell \nu}^2}d {q_{d_j \nu}^2}} } \label{eq:assymdef}
\end{align}
where the barred cross section $\bar\sigma$ refers to the cross section of the CP-conjugate interaction. Note that because we are working with $2\to 3$ body interactions utilizing baryonic initial states, we cannot write a simple analytic expression for this asymmetry. For the purpose of assessing the size and variation of the asymmetry with model parameters, we approximate the cross-sections as $1\to 3$ processes to capture the pole effects, and neglect phase-space assuming that these effects can be factorized from the numerator and denominator of the asymmetry in Eq.~\eqref{eq:assymdef}. To circumvent this assumption, in Sec.~\ref{sec:analysis} we perform numerical studies on asymmetry as a $2\to 3$ process incorporating particle-level effects.  Below we will discuss weak and strong phase interference in the context of an EFT parameterization in Sec.~\ref{sec:EFTweakstrong} and in a concrete simple UV completion in  Sec.~\ref{sec:BSMweakstrong}.

\subsubsection{\texorpdfstring{Treatment with an EFT parameterization of BSM}{Weak and strong CP phases in ppdellnu with EFT}}
\label{sec:EFTweakstrong}
Assuming the BSM interactions are generated at sufficiently high energy, the $W$ pole can be utilized to probe effective interactions for which the BSM dynamics cannot be resolved. The relevant strong phase which will enter into this study will be that of this $W$ propagator, given by the Breit-Wigner functional form for massive vector bosons,
\begin{align}
    i \Sigma _{\mu\nu}(q) = -i \frac{g_{\mu\nu}- q_\mu q_\nu/M^2}{q^2-M^2+iM\Gamma}\,.
    \label{eq:BWdist}
\end{align}
In the above equation, $q$ is the momentum carried by the propagator, $M$ is the propagator mass, and $\Gamma$ is the full decay width of the propagator. We denote  $(q^{\ell\nu})^2$ to represent the invariant mass of the $\ell-\nu_\ell$ pair in the final state. When the propagator illustrated in Fig.~\ref{fig:pptaunu} is on-shell, $q^2= (q^{\ell\nu})^2=M^2$, and 
\begin{align}
    i \Sigma _{\mu\nu}^{\text{on-shell}}(q_{\ell\nu}) = \frac{g_{\mu\nu}- (q_{\ell\nu})_\mu (q_{\ell\nu})_\nu/m_W^2}{m_W\Gamma_W}\,,
\end{align}
where $m_W$ is the $W$ boson mass, and $\Gamma_W$ is the $W$ boson full decay width.
By parameterizing the effective BSM interaction with the dimensionless coupling coefficient $C_{V_L}$, one can see that the total amplitude of this process will scale as 
\begin{align}
    i \mathcal{M} = i (\mathcal{M}_{\text{BSM}}+\mathcal{M}_{\text{SM}})
    &\propto  -2\sqrt{2} G_F V_{ij}^*\left(\frac{ m_W }{{ \Gamma_W}} + i C_{V_L}^{ji \ell, *}\right)\,,
\end{align}
where $G_F$ is the Fermi constant.  Therefore, the differential cross section will scale with the square of the amplitude as
\begin{align}
\label{eq:diffxsecEFT}
   \frac{d \hat\sigma}{d q_{\ell \nu}^2 q_{d_j\nu}^2}
   &\propto 8\,G_F^2 |V_{ij}|^2 \times \left(\frac{m_W^2}{\Gamma_W^2} +|C_{V_L}^{ji  \ell}|^2-\frac{2m_W}{\Gamma_W V_{ij}}\text{Im}(V_{ij}C_{V_L}^{ji  \ell})\right)\,, 
\end{align}
where $q_{d_j\nu}$ is the invariant mass of the $d_j\nu$ in the final state.

Close to the $W$ pole, the asymmetry in Eq.~\eqref{eq:assymdef} is
\begin{align}
\label{eq:Wassym}
      \mathcal{A}_\text{CP}|_\text{W cuts} \approx &\frac{ 2\hat{m}_W \text{Re}(V_{ij})\text{Im}(V_{ij} C_{V_L}^{ji \ell})}{|V_{ij}|^2 (\hat{m}_W^2 +|C_{V_L }^{ji \ell}|^2)}\,,\\
    &= \frac{ 2\hat{m}_W \text{Re}(V_{ij})|C^{ji\ell}_{V_L}| }{|V_{ij}|^2 (\hat{m}_W^2 +|C_{V_L }^{ji \ell}|^2)} \left( \text{Re}(V_{ij}) \sin \phi + \text{Im}(V_{ij}) \cos \phi \right)
    \label{eq:Wassym1}
\end{align}
with $\hat{m}_W= m_W/\Gamma_W$,  $\text{Im}(C^{ji\ell}_{V_L})= |C^{ji\ell}_{V_L}| \sin \phi$ and $\text{Re}(C^{ji\ell}_{V_L})= |C^{ji\ell}_{V_L}| \cos \phi$. Note that the expression in Eq.~\eqref{eq:Wassym} does \emph{not} assume that the CKM element is real, and so the nonzero SM CKM phase will, in principle, contribute to the asymmetry even if the BSM introduces no additional phase. 

Ideally, one could apply kinematic cuts to $q_{\ell\nu}$ to isolate the region where there is maximal impact of this phase-interference. However, the neutrino itself cannot be detected in a collider. This hurdle can be overcome  by considering observables that encode information about the invariant mass, such as the transverse mass of the $d_j+$MET and  $\ell+$MET. This will be further discussed in the context of a numerical analysis in Sec.~\ref{sec:analysis}. Notwithstanding, the $W$ pole represents a region of phase space with large SM backgrounds, making precision studies in this region exceptionally difficult. This motivates considering the impact of an additional CP-even phase, via some BSM propagator. 

\begin{figure}
    \centering
\includegraphics[width=0.4\linewidth]{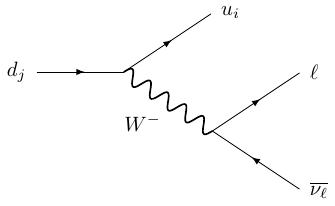}\hspace{5mm}
\includegraphics[width=0.4\linewidth]{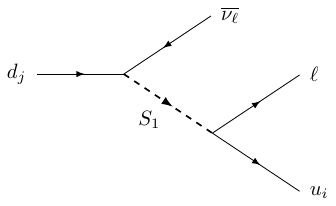}
   \caption{Tree-level contribution to $d_j\to u_i \ell\nu_\ell$ from the SM $W$ (left) and the $S_1$ leptoquark (right).}
    \label{fig:bctaunu}
\end{figure}

\subsubsection{\texorpdfstring{A toy model featuring an isosinglet scalar leptoquark}{Weak and strong CP phases in ppdellnu with prop}}
\label{sec:BSMweakstrong}

Consider now the addition of a scalar which can mediate the interactions in Fig.~\ref{fig:bctaunu}. For now, we remain agnostic to how the effective coupling $C_{V_L}$ is generated (this will be discussed in Sec.~\ref{sec:BSMweakstrong}).
Instead, here we focus on that there is permutation of the final state particles with respect to the mediator when contrasted with the SM process, see Fig.~\ref{fig:bctaunu}. Our mediator of choice is the scalar LQ $S_1 \sim (\mathbf{3}, 1, -1/3)$\footnote{Where the entries of the tuple here refer to the transformation properties under the SM Gauge group $SU(3)\otimes SU(2)\otimes U(1)$.}. Scalar LQs provide a portal for generating direct quark-lepton interactions. Note that LQs are particularly appealing for studies such as these because their direct lepton-quark couplings naturally permute the final state particles relative to the internal propagator. Therefore, the two invariant mass combinations in the final state may be exploited to seek the effects of distinct propagator phases on the differential cross-section. 

In $pp[u_i g] \to d_j \ell \nu_\ell$, one would expect resonance behavior around $q_{d_j \nu}^2\sim M_{S_1}^2$ which sources an additional CP-even phase generated by the $S_1$ propagator. The utilization of both propagators with differing  final state momentum topologies is the handle with which one can define the a contribution to a charge asymmetry proportional to a CP-violating observable akin to that in reference~\cite{Berger:2011wh}.

Working at energies above the electroweak scale, $q_{\ell\nu}^2 \gg m_W^2$ and around $M_{S_1}$, the EFT becomes invalid and one should consider the LQ as a propagating field rather than integrating it out to form an EFT. In this region, the amplitude representing the interactions in Fig.~\ref{fig:bctaunu} takes the form 
\begin{align}
    i\mathcal{M} \propto -i 2\sqrt{2}G_F V_{ij}^* \left(\frac{(C^{ji\ell}_{V_L})^*M_{S_1}^2}{q_{d_j \nu}^2-M_{S_1}^2+iM_{S_1}\Gamma_{S_1}} +\frac{ m_W^2 }{q_{\ell\nu}^2} \right)\,,
\end{align}
where we have employed a Breit-Wigner distribution for a propagating scalar, which has a similar form to Eq.~\eqref{eq:BWdist} but without Lorentz structure. Close to the LQ mass scale in the $d_j-\nu$ momentum distribution, $(q_{d_j\nu_j})^2 \sim M_{S_1}^2$, the amplitude becomes
\begin{align}
    i\mathcal{M} \propto -i 2\sqrt{2}G_F V_{ij}^* \left(-i(C^{ji\ell}_{V_L})^*\frac{M_{S_1}}{\Gamma_{S_1}} +\frac{ m_W^2 }{q_{\ell\nu}^2} \right)\,.
\end{align}
 Assuming for brevity that we may work in a basis where the CKM element $V_{ij}$ is real, 
\begin{align}
    \frac{d\hat\sigma}{d {q_{\ell \nu}^2}d {q_{d_j \nu}^2}} &\propto 8G_F^2 |V_{ij}|^2  \times\left(|C^{ji\ell}_{V_L}|^2\frac{M_{S_1}^2}{\Gamma_{S_1}^2} 
    - 2 \frac{m_W^2 M_{S_1}}{q_{\ell\nu}^2 \Gamma_{S_1}}\text{Im}(C^{ji\ell}_{V_L})
    +\frac{ m_W^4 }{q_{\ell\nu}^4} \right) \,,
    \label{eq:diffxsec}
\end{align}
so we have that the pole of the LQ propagator isolates the imaginary component of the WC. Around the LQ pole, the tree-level partonic asymmetry will have the form 
\begin{align}
      \mathcal{A}_\text{CP}|_\text{LQ cuts} \approx -\frac{ 2\hat{M}_{S_1} \frac{m_W^2 }{(q^{\ell\nu})^2}|C^{ji\ell}_{V_L}| \sin \theta_{S_1}}{\frac{ m_W^4 }{(q^{\ell\nu})^4}+|C^{ji\ell}_{V_L}|^2 \hat{M}_{S_1}^2 }. \label{eq:LQassym}
\end{align}
where we fix $\hat{M}_{S_1}\equiv \frac{M_{S_1}}{\Gamma_{S_1}}$. 

The behavior of the asymmetry around both the $W$, see Eq.~\eqref{eq:Wassym}, and the scalar $S_1$, see Eq.~\eqref{eq:LQassym},  represents two cases of limiting behavior of the differential cross sections $ {d\hat\sigma}/{d {q_{\ell \nu}^2}d {q_{d_j \nu}^2}}$. Moving away from each of these poles, the  interference effects vary over the $({q_{\ell \nu}^2},{q_{d_j \nu}^2})$ plane. One can then imagine examining the anisotropy of differential cross section over this plane to search for hints of CP-violation, akin to how a Dalitz plot is utilized for $1\to 3$ body decays. Note also that the sign of the asymmetry is expected to flip as the dominant CP-even phase moves from one pole to another.  

It remains left to discuss the viability for extracting such an asymmetry from $pp$ collisions, and also the relative size of this asymmetry in different kinematic regions. 
However, we will first remark on the utilization of such a result. Firstly, this has the potential to be a high-energy observable effect of CP-violation, visible in colliders beyond the context of meson decays and oscillation. Moreover, it highlights the importance of considering models beyond the EFT in some probes of CP-violation in colliders.

We will now aim to provide a proof-of-principle of the potential for studying this asymmetry at colliders utilizing a toy model.  We focus here on the application of BSM which too can modify charged-current semileptonic $B$ decays.

\subsection{\texorpdfstring{Application to $u_i g\to b \tau \nu_\ell$}{Application to ppbtaunu}}
\label{sec:cpv_bsm}

 There is a long standing anomaly in the ratio the charged-current $B$ decays to different charged-lepton final states
$R_{D^{(*)}}={\text{BR}(B\to D^{(*)} \tau\nu)}/{\text{BR}(B\to D^{(*)} \ell\nu)}\,,$
where $\ell\in \{ e, \mu\}$, a precision test of lepton-flavor universality. The result of a
fit to experimental data by the Heavy Flavor Averaging Group~(HFLAV)~\cite{HFLAV}\footnote{HFLAV~\cite{HFLAV} combines the experimental results of \cite{BaBar:2012obs,BaBar:2013mob,Belle:2019rba,LHCb:2023cjr,Belle:2015qfa,Belle:2016dyj,Belle:2017ilt} with
the theory predictions of \cite{Bigi:2016mdz,Gambino:2019sif,Bordone:2019vic,Bernlochner:2017jka,Jaiswal:2017rve,BaBar:2019vpl,Martinelli:2021onb}.} leads to larger values for $R(D)$ and $R(D^*)$ 
than expected and exhibits a tension with the SM
prediction at the $\sim 3 \sigma$ level.
This tension has led many in the BSM community to construct models which modify the charged-current interaction $b\to c \tau \nu$, to enhance this final state over the decay to lighter flavors  (see e.g.~\cite{ Aebischer:2022oqe,Iguro:2023prq,Becirevic:2024pni}). By crossing symmetry, this process is related to the Drell-Yan process $pp [bc]\to \tau\nu$, and to 
$pp [g c]\to b \tau\nu$, where the square brackets denote the quark-level structure of the interactions. Thus an enhancement in this meson decay at the $\mu_B=4.2$ GeV (low-energy, $B$-meson scale) can have effects that can be searched for and constrained by these collider processes (probed at $\mu_{pp}\sim$ TeV). 

Several papers have focused on using high-$p_T$ tails of kinematic distributions in $pp$ collisions to constrain BSM, from $pp\to \ell\nu$~\cite{Angelescu:2020uug, Greljo:2018tzh,Fuentes-Martin:2020lea,Allwicher:2022gkm} and $pp\to b\ell\nu$~\cite{Abdullah:2018ets,Marzocca:2020ueu,Endo:2021lhi}. There is a heightened interest in the third-generation final states as these are the least-well experimentally constrained. In order to probe these final-states, we require techniques like $b$-tagging and $\tau$-tagging to control the significant backgrounds. The advent of increased sensitivity in $\tau$-tagging and $b$-jet reconstruction makes accurate measurements of these final-states more feasible. 

In the existing studies outlined above, the impact of imaginary components to the associated WCs was not considered as they largely focused on a kinematic region where they deemed inteference to be negligible. A detailed study of the influence of BSM-SM phase-interference in $pp\to b \tau\nu$ is, to our knowledge, not yet present in the literature. 

One of the simplest models for explaining a deviation from the SM in charged-current decays are based on scalar leptoquark extensions to the SM. Scalar leptoquarks provide a portal for generating direct quark-lepton interactions. Notably, the isosinglet leptoquark $S_1$, plays a key role in many extensions of the SM that aim to explain the deviation from the SM in $R_{D^{(*)}}$ by enhancing the contribution to $b\to c\tau\nu$ via the tree-level diagram shown in Fig.~\ref{fig:bctaunu}.\footnote{For an excellent review of the physics of scalar leptoquarks, see reference~\cite{Dorsner:2016wpm}.}

The interactions of this LQ with SM fermions are given by the following, after EWSB, 
\begin{equation}
\begin{aligned}
\mathcal{L}_{S_1} 
\supset \lambda_L^{ij} &\left[\overline{\nu^c_i} P_L d_j  - V^\dagger_{jk} \overline{e^c_i} P_L u_k \right]S_1^*  + \lambda_R^{ij} \overline{e^c_i} P_R u_j S_1^* + \text{h.c.} 
\end{aligned}
\end{equation}
working in the down-type-quark mass diagonal basis (see Appendix B).\footnote{We chose  the order of the indices in $\lambda^{ij}$ such that $i$ refers to lepton flavour, then $j$ to quark flavour, so that leptoquark can act
as a mnemonic.} To address the anomalies in $b\to c \tau\nu$, a minimal coupling setup consists of fixing $(\lambda_{L}^{33},\lambda_{L}^{32}, \lambda_{R}^{32})$ to nonzero values, therefore generating the following effective interactions (corresponding to dimension-6 operators in the WET EFT basis, as defined in Eq.~\eqref{eq:EFTbasis}),
\begin{align}
C^{S_1}_{S_L} (M_{S_1}) &= -4 C_{T}(M_{S_1}) = \frac{(\lambda_R^{32})^* \lambda_L^{33}}{4\sqrt{2}G_FV_{cb} M_{S_1}^2} ,\\
C_{V_L, i}^{S_1}(M_{S_1}) 
&= \frac{\lambda_L^{33}\left[(\lambda_L^{33})^* V_{ib}+(\lambda_L^{32})^*V_{is}\right]}{4\sqrt{2}G_FV_{ib} M_{S_1}^2} 
\label{eq:CVL_def}
\end{align}
where $C_{V_L, i}^{S_1}$ is the generated effective coupling containing the up-type quark $u_i$ (vector couplings to all up-type quarks are generated in this framework via the CKM).  The above are taken to be algebraic relations fixed at the $M_{S_1}$ energy scale.

For the purpose of addressing the flavour anomalies, one requires the presence of both the scalar/tensor and vector WC in order to align with present constraints (see e.g.\cite{Becirevic:2024pni}). However, here our focus will not be on presenting a model that can address the anomalies, but rather studying the impact of new physics resonances in charged-current decays. For our study below, we seek a BSM effect that interferes with the SM and therefore focus-in on $C_{V_L}^{S_1}\neq 0$.
Practically, this may be achieved by fixing $\lambda^{32}_R=0$ and varying the other relevant couplings. The presence of additional Lorentz structures will enlarge the BSM-only term, and is expected to reduce the size of the observable asymmetry. 

For studying any interference effect due to the LQ propagator, there will be dependence on $\Gamma_{S_1}$. If the LQ decays are prompt, corresponding naively to large LQ couplings, then the width is large and one would hope to amplify interference effects over a broader kinematic region. Here we do not presume that the LQ decays solely to the particles considered in the decay process outlined above, and treat the LQ width as a free-parameter which obtains contributions from other decay pathways. For the purpose of understanding this effect, we assume the physics modifying the decay width of the LQ has no bearing on the interactions relevant for the processes studied here.

So far in this work, the analytic treatment of the cross-section and asymmetry scaling has been at the parton level and not considered the phase space of the final state particles or the initial-state gluon interactions. Firstly, phase space distinction between the $2\to 3$ and $1\to 3$ decay will largely factorize from the numerator and denominator, but other intermediate states will distort the assumptions utilized in defining the expression in Eq.~\eqref{eq:LQassym}, but not in establishing the presence of the asymmetry. In order to correctly parameterize these effects one requires a careful (numerical) treatment of the differential cross-sections which will be done below. Secondly and more pressingly, the partonic cross-section used in the above analytics neglects the impact of proton Parton Distribution Functions~(PDFs), parton showering, hadronization, handling the neutrino, and tagging efficiencies on the expected signal yield. Throughout this study, we make distinction of partonic event as $u_i g\to b \tau \nu$, parton-level as $p\,p\to b \tau \nu$ without any parton shower or hadronization, and particle-level as $p\,p\to b \tau \nu$ with parton showers, hadronization and detector smearing.
The impact of these are discussed and addressed below in our numerical study.

\section{Numerical study}
\label{sec:analysis}

Existing strategies for probing charged-current interactions in $pp$ collisions are only able to set limits on the magnitude of the WCs (see e.g. Refs.~\cite{Abdullah:2018ets,Marzocca:2020ueu,Endo:2021lhi}). In this section, we develop a methodology for extracting the magnitude and phase of the left-handed vector interactions utilizing the novel observable outlined above.
Here we do not attempt to isolate the effects of this interference near either the $W$-pole or the LQ-pole (for which an EFT-driven study is insufficient), but instead look at the entire phase space region and investigate the limitations of this method. Moreover, by not specifying a search around a particular BSM pole leaves the broad methods of this numerical study agnostic to the nature of the model. 

To conduct an initial analysis of the asymmetry, it is necessary to express it in terms physical quantities reconstructed in collider experiments. That is, in terms of the transverse mass of $b+$MET and $\tau+$MET, where MET is the missing transverse energy used as a proxy for the neutrino,
as neutrinos are invisible to the LHC detectors. This is defined as
\begin{equation}
M_T^{a\nu} = \sqrt{2p_T^ap_T^\nu (1-\cos\theta)}\,,
\label{eq:mt}
\end{equation}
where $a$ represents the visible particle, and $\theta$ is the angle between the particle and the missing transverse momentum in the transverse plane. This observable is extensively used when trying to understand the mass of particles that contain a single neutrino in the final state, e.g. the $W$ mass measurements (see for example Refs.~\cite{CDF:2022hxs,Isaacson:2022rts,ATLAS:2024erm}).  As an experimental proxy for Eq.~\eqref{eq:assymdef}, we propose the use of
\begin{align}
    \mathcal{A}_\pm^\text{exp} = \frac{{dN}/{d {M_T^{\tau \nu}}d {M_T^{b \nu}}}- {d\bar N}/{d {M_T^{\tau \nu}}d {M_T^{b \nu}}}  }{{dN}/{d {M_T^{\tau \nu}}d {M_T^{b \nu}}}+{d\bar N}/{d {M_T^{\tau \nu}}d {M_T^{b \nu}}}} , \label{eq:expassymdef}
\end{align}
\noindent where ${dN}/{d {M_T^{\tau \nu}}d {M_T^{b \nu}}}$ is the number of events ($N$) as a function of transverse mass among final states. 
During this analysis, we calculate the transverse mass for both the $\tau-\nu$ and $b-\nu$ systems, classified as the ``plus'' and ``minus'' states aligning with the charge of the $\tau$-jet.

In contrast to the analytic study, these experimental observables are sensitive only to transverse momentum, discarding information about the final-state momentum in the longitudinal direction.
This will have the effect of smearing and (potentially) reducing the measured asymmetry in the observable kinematic region.
This information loss is particularly relevant for observables utilizing MET to parameterize the neutrino especially in the case of multiple neutrinos within an event (e.g. from the decay of the tau).
Furthermore, precise measurement of the MET is challenging due to jet energy reconstruction uncertainties, pileup, and gaps within the detector -- see Ref.~\cite{ATLAS:2018txj} for a detailed study by ATLAS. We investigate the performance of Eq.~\ref{eq:expassymdef} parameterized in terms of the MET, at parton level, in Section~\ref{ssec:valid} , and at particle level in Section~\ref{sec:particle_level}.

To validate the selection of $\mathcal{A}_\pm^\text{exp}$ as the measurable quantity for characterizing the asymmetry in Eq.~\eqref{eq:assymdef}, an initial partonic analysis is conducted to study the variation of asymmetry with the phase $\theta_{S_1}$ in Sec.~\ref{ssec:valid}, and a further study of variation of the asymmetry over the two-dimensional transverse-momentum plane in  Sec.~\ref{ssec:kinem}. This is followed by a particle-level analysis, incorporating a rough detector smearing, to estimate the experimental sensitivity for a proposed search in Sec.~\ref{sec:particle_level}. However, before proceeding with the analysis we will address the backgrounds to each study below.

\subsection{Backgrounds}
\label{sec:bg}
For the partonic analysis (Sec.~\ref{sec:partonlevel}), we only consider the irreducible background ($W+b$ jet), since we can easily remove other processes at truth level by requiring exactly one $\tau$ and one $b$ quark. At the particle-level (Sec.~\ref{sec:particle_level}), additional backgrounds arise from tagging efficiencies, smearing effects, etc; the dominant backgrounds are $Z$+jets, $t\bar{t}$+jets, and $W$+light jets (where `light jets' refers to anything but a $b$ jet).

In the particle-level study, we include all the below backgrounds. For the $Z$+jets background, the final state would consist of the $Z$ decaying to a pair of tau leptons.
In this case, the efficiency in tagging the $\tau$-jets is symmetrical under the electric charge of the tau.
Therefore, any $ \mathcal{A}_\pm^\text{exp}$ in this background should be statistically consistent with zero, and the choice of cuts can be optimized to ensure that this background is negligible.
Similarly, the $t\bar{t}$+jets background is symmetric in the final state, even when incorporating  decays of only a single top. 
Thus, the expected $ \mathcal{A}_\pm^\text{exp}$ due to this background should also be statistically consistent with zero.
Finally, we include the $W$+light jets, which provides the dominant background to the signal process due to its large cross section and the high miss-tagging rate for, especially, $c$ jets.

\begin{figure}
    \centering
    \includegraphics[width=0.5\textwidth]{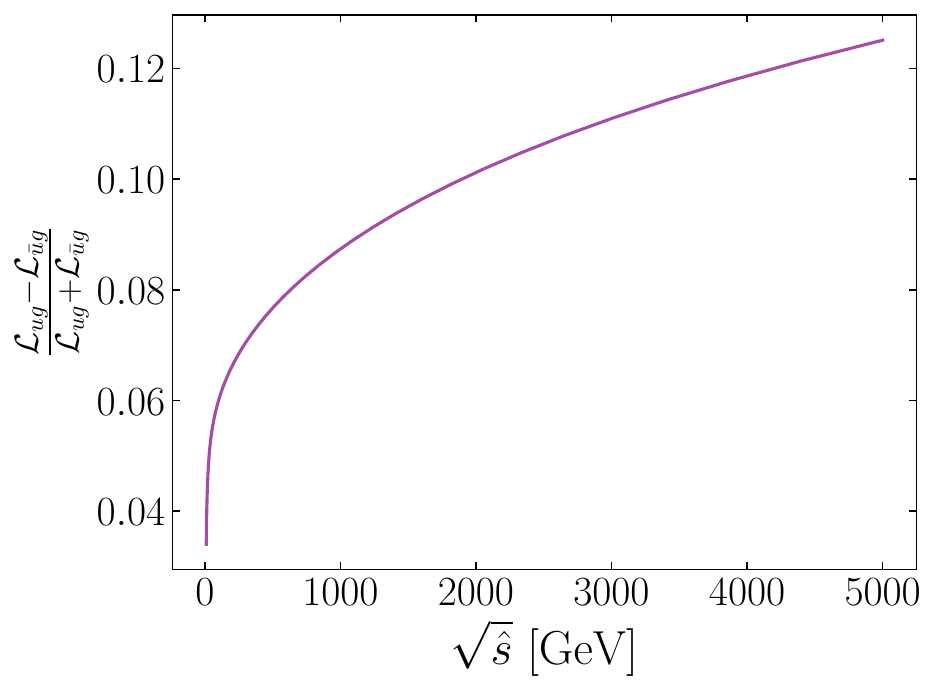}
    \caption{The luminosity asymmetry between the $ug$ and $\bar{u}g$ initial states.}
    \label{fig:lumi_asym0}
\end{figure}

Notably, LHC collisions occur between two proton beams and therefore PDFs are incorporated when considering the collisions at particle-level. The $c$ and $\bar{c}$ content of a proton are expected to be equal as they are both sea quarks. This makes a $cg$ initial state from proton collisions an appealing target for an asymmetry measurement, as the collision of two protons is symmetric in this initial state. However, the $ug$ initial state, containing a valence $u$ quark, is an irreducible background when comparing the rates of $pp\to b \tau^+ \nu$ to $pp\to \bar{b} \tau^- \bar{\nu}$. Although this subprocess is CKM suppressed in the SM, as $|V_{ub}|\ll |V_{cb}|$, the baseline luminosity of $u$ and $\bar{u}$ from working with protons and not antiprotons is unavoidable. This luminosity contribution to this asymmetry is illustrated in Fig.~\ref{fig:lumi_asym0}. For low partonic center-of-mass energy,  $\sqrt{\hat{s}}$, the asymmetry is small, but it increases with $\sqrt{\hat{s}}$. As the maximal asymmetries occur around the location of the $W$ or BSM pole, the PDF asymmetry will be larger around the pole of a LQ with a larger mass. Nevertheless, we can predict the background asymmetry from the PDFs and this may be subtracted from the observed asymmetry in a full analysis. 

\subsection{Parton-level analysis}
\label{sec:partonlevel}
We begin by performing a parton-level $pp \to b\tau\nu_\tau$ analysis for the LHC setup.
Events are simulated using \texttt{Sherpa} version 3.0 with the \texttt{COMIX} matrix element generator~\cite{Gleisberg:2008fv,Hoche:2014kca,Bothmann:2024wqs}. The parton distribution functions are given by \texttt{NNPDF3.0}~\cite{Hartland_2013, Ball_2015} via the use of \texttt{LHAPDF6}~\cite{Buckley:2014ana}. The signal events may be divided into three categories, depending on BSM expansion order ($\mathcal{O}(\mathcal{M}_\text{BSM}^n)$) of $|\mathcal{M}|^2$ as shown in Eq.~\eqref{eq:BSMSM}. These categories are: interference only ($n=1$), SM $+$ interference ($n \leq 1$) and the full expansion ($ n\leq 2$). The first gives a small contribution and requires more robust techniques to isolate its effect. For this analysis, and henceforth, we consider the full expansion as our signal events. 

To generate signal events, we utilize a \texttt{UFO} model file~\cite{Degrande_2012,Darme:2023jdn}, including the full SM Lagrangian with a non-diagonal and (generally) complex CKM matrix, and the LQ $S_1$~\cite{Dorsner:2018ynv}~\footnote{The model file can be accessed at~\cite{bigaran_2024_12576982}.}. The complex-valued couplings are varied by adjusting the radial and phase components of $\lambda_L^{32}$, while keeping the other LQ coupling parameters fixed at real values. Under this convention, and utilizing the expressions in Sec.~\ref{sec:BSMweakstrong}, we can link weak CP-phase dependencies to the asymmetry observed. In particular, to enhance and account for the $(\lambda_L^{32})^*V_{is}$ contribution in Eq.~\eqref{eq:CVL_def}, we consider throughout this  analysis a signal model with real-valued CKM (i.e. $V_{ij} \in \mathbb{R}$),
$\lambda_R^{32} = 0 $ and $\lambda_L^{33} = 1$.
We employ a LQ mass of 1.5 TeV, which is just on the boundary of current limits~\cite{ATLAS:2023uox}\footnote{Note that in a concrete model, these direct limits should be reinterpreted incorporating also the LQ self-corrections from~\cite{Bigaran:2024vnl}.}. 
Our benchmark is chosen to have $|C_{V_L}|=1$, which
fixes the value of $\lambda_L^{32}$ as a function of the phase of $C_{V_L}$ ($\theta_{S_1}$) using Eq.~\eqref{eq:CVL_def}. The decay width of LQ is set to around 1\% of the LQ mass, consistent with the coupling assignment, which can draw out full benefits from the narrow width approximation as described in Eq.~\eqref{eq:diffxsec}.
For the parton-level study, we only consider the irreducible background that arises from $W+b$ jet events, as discussed in Sec.~\ref{sec:bg}.

\begin{table*}[t]
\begin{centering}
    \begin{tabular}{|c|c|}
        \hline
      Observable & Cut\\
        \hline\hline
        $p_T(b)$  & $>20$ GeV\\
        $p_T(\tau)$  & $>20$ GeV\\
        MET & $>20$ GeV\\
        \hline 
        \end{tabular}
        \caption{Minimial transverse-momentum, generator-level cuts applied to event simulation.}
         \label{tab:MinimalCuts}
        \end{centering}
\end{table*}
    
The parton level events are generated with minimal transverse momentum cuts, listed in Tab.~\ref{tab:MinimalCuts}.
These cuts are chosen to maximize the cross-section, although traditionally would inhibit disentangling signal from the background.

\subsubsection{Validating the experimental asymmetry}
\label{ssec:valid}

\begin{figure}[t!]
    \centering   \includegraphics[width=0.45\linewidth]{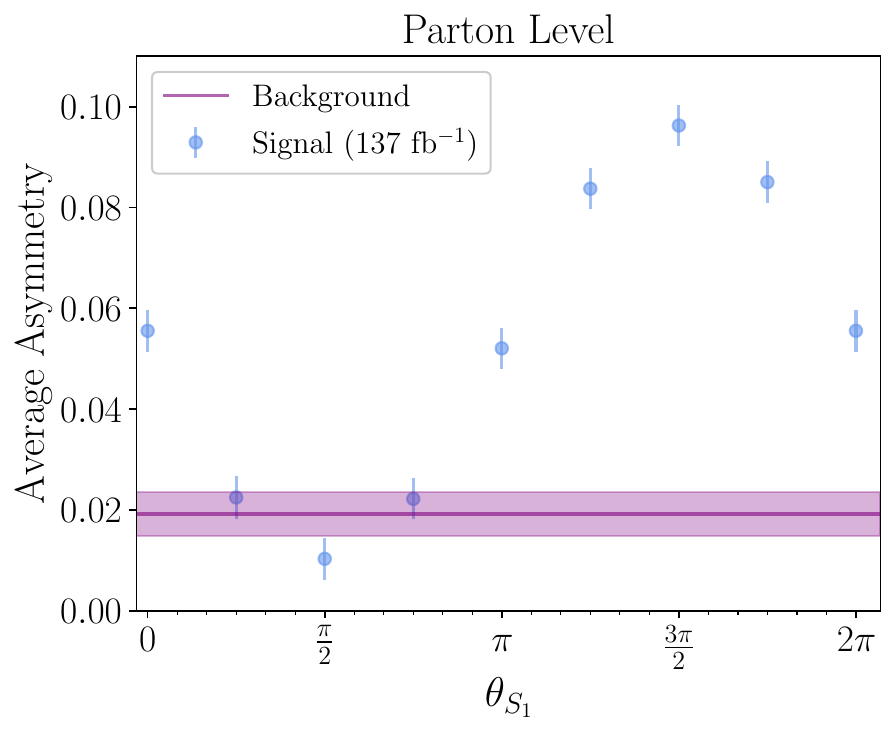}\hspace{1cm}
\includegraphics[width=0.45\linewidth]{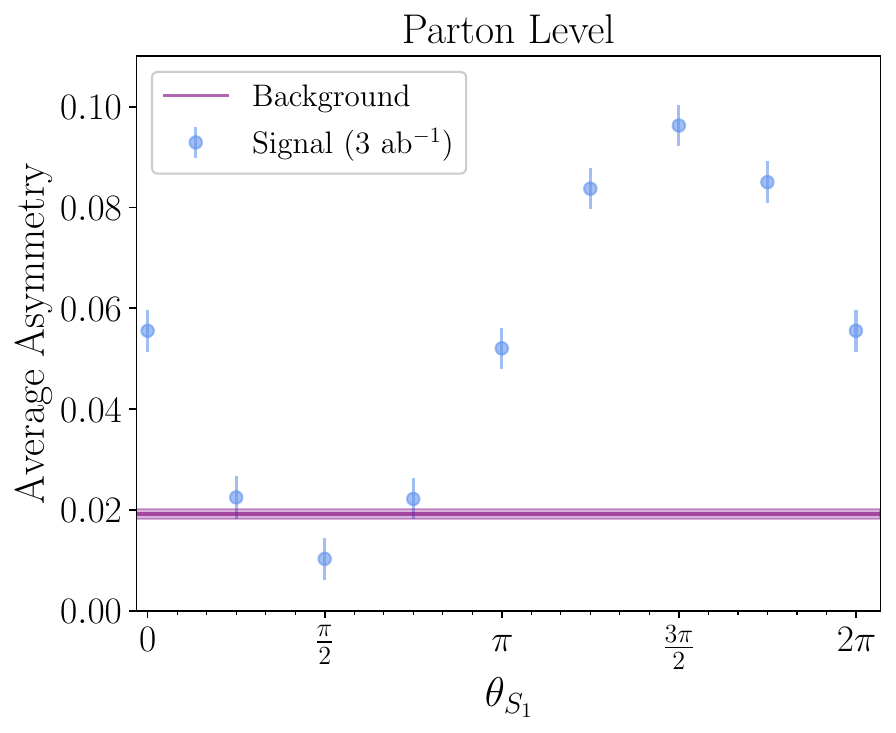}
    \caption{Average asymmetry of the $pp \to b\tau\nu_\tau$ signal and $W$+jets background for different values of the $C_{V_L}$ phase ($\theta_{S_1}$). In both cases the signal is simulated at parton level for a LQ mass of $M_S = 1.5$ TeV,  $|C_{V_L}|=1$, and minimal transverse momentum cuts, listed in Tab.~\ref{tab:MinimalCuts}. In the right panel the signal is generated for a luminosity of $137$ fb$^{-1}$ and in the left panel for a projected luminosity for the HL-LHC of $3$ ab$^{-1}$. The uncertainty bands represent the statistical uncertainties for a given luminosity. }
    \label{fig:SigandBG}
\end{figure}

We begin by validating the variation of $\mathcal{A}_\pm^\text{exp}$ with the BSM phase, $\theta_{S_1}$ to assess whether the use of transverse momenta washes out the effect. The signal is simulated at parton level for a LQ mass of $M_S = 1.5$ TeV and fixing $|C_{V_L}| = 1$. We scan over $\theta_{S_1}$ between zero and $2\pi$, and calculate the average value of $\mathcal{A}_\pm^\text{exp}$ over the two-dimensional kinematic distributions in $M_T^{\tau\nu}$ and $M_T^{b\nu}$. Both signal and background have minimal transverse momentum cuts listed in Tab.~\ref{tab:MinimalCuts} applied at generator level. Fig.~\ref{fig:SigandBG} illustrates the distribution of this average asymmetry for a luminosity of $137$ fb$^{-1}$ (right panel) and for projected luminosity for the HL-LHC of $3$ ab$^{-1}$ (left panel).
The uncertainty bands in the figure are given solely from the statistical uncertainty expected at $137$ fb${}^{-1}$ and $3$ ab${}^{-1}$ respectively.
Since the observable is an asymmetry, the systematic uncertainty is expected to be drastically reduced, but a detailed study of the systematic uncertainty is beyond the scope of this work.

Recall that the full theoretical asymmetry is given by Eq.~\eqref{eq:assymdef}, with limiting behaviour around the LQ and W poles given by Eq.~\eqref{eq:LQassym} and Eq.~\eqref{eq:Wassym1}, respectively. We did not provide a full analytic expression for the asymmetry under the theoretical assumptions, as the experimental limitations anyway would require a definition of a proxy observable. Contrasting the results for $\mathcal{A}_\pm^\text{exp}$ in Fig.~\ref{fig:SigandBG} to the limiting cases of Eq.~\eqref{eq:LQassym} and Eq.~\eqref{eq:Wassym1}, note that the same oscillatory behavior with $\theta_{S_1}$ is observed, recalling that the CKM phase in the signal is taken to be negligible. The expected minimum at $\theta_{S_1}=\pi/2$ and maximum at $\theta_{S_1}=3\pi/2$ are observed, demonstrating that the pole-effects are, indeed, dominant in dictating the behavior of this asymmetry. The asymmetry, however, does not oscillate around zero: even when the phase is fixed to zero, the background asymmetry is largely generated by the proton's $u-\bar u$ PDF asymmetry in $pp\to b \tau \nu$ -- see Fig.~\ref{fig:lumi_asym0}, smeared by the transverse-mass reconstruction. Recall that in the background $W+b$ generation, we adopt a complex CKM and so in addition to the PDF asymmetry for this process, there is also a contribution from the CKM phase -- see e.g. Eq.~\ref{eq:Wassym1}. Note that this asymmetry is averaged over all accessible phase space, and could be enhanced by cutting around a particular BSM or SM pole.

The separation between signal and background is not uniform for all $\theta_{S_1}$ phases. However, with increased statistical precision at the HL-LHC, this separation becomes notably more distinct. This too is true for a particle-level analysis, as we will see in Section~\ref{sec:particle_level}.

\begin{figure}[t!]
    \includegraphics[width=\textwidth]{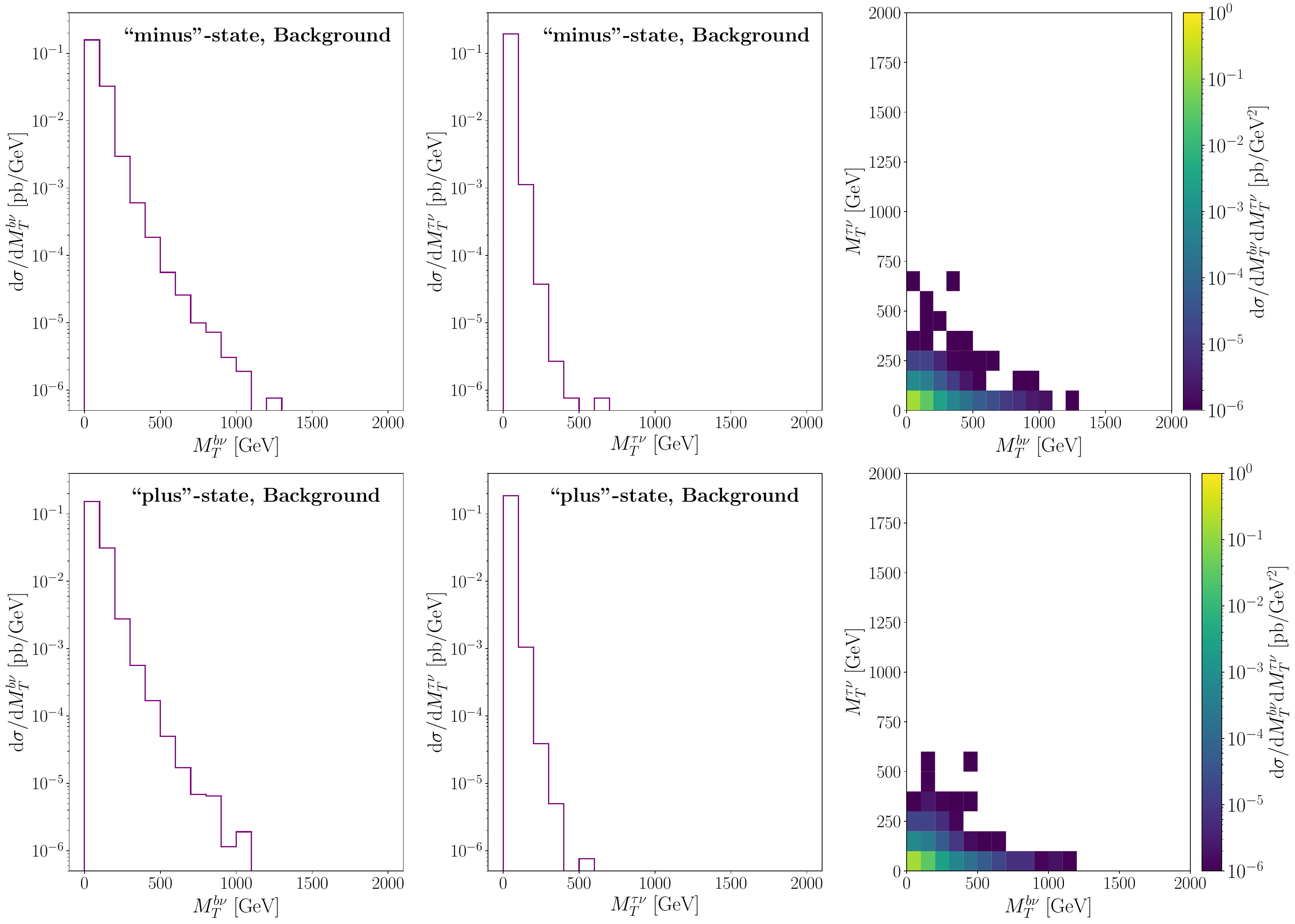}
    \caption{A comparison of background events from the ``minus''-state (top) and the ``plus''-state (bottom) for the transverse mass of the $b-\nu$ system on the left, $\tau-\nu$ system in the middle, and the 2D distribution of the transverse masses on the right.}
    \label{fig:mt_bkg}
\end{figure}

\subsubsection{Two-dimensional kinematics}
\label{ssec:kinem}

\begin{figure}[t!]
    \includegraphics[width=\textwidth]{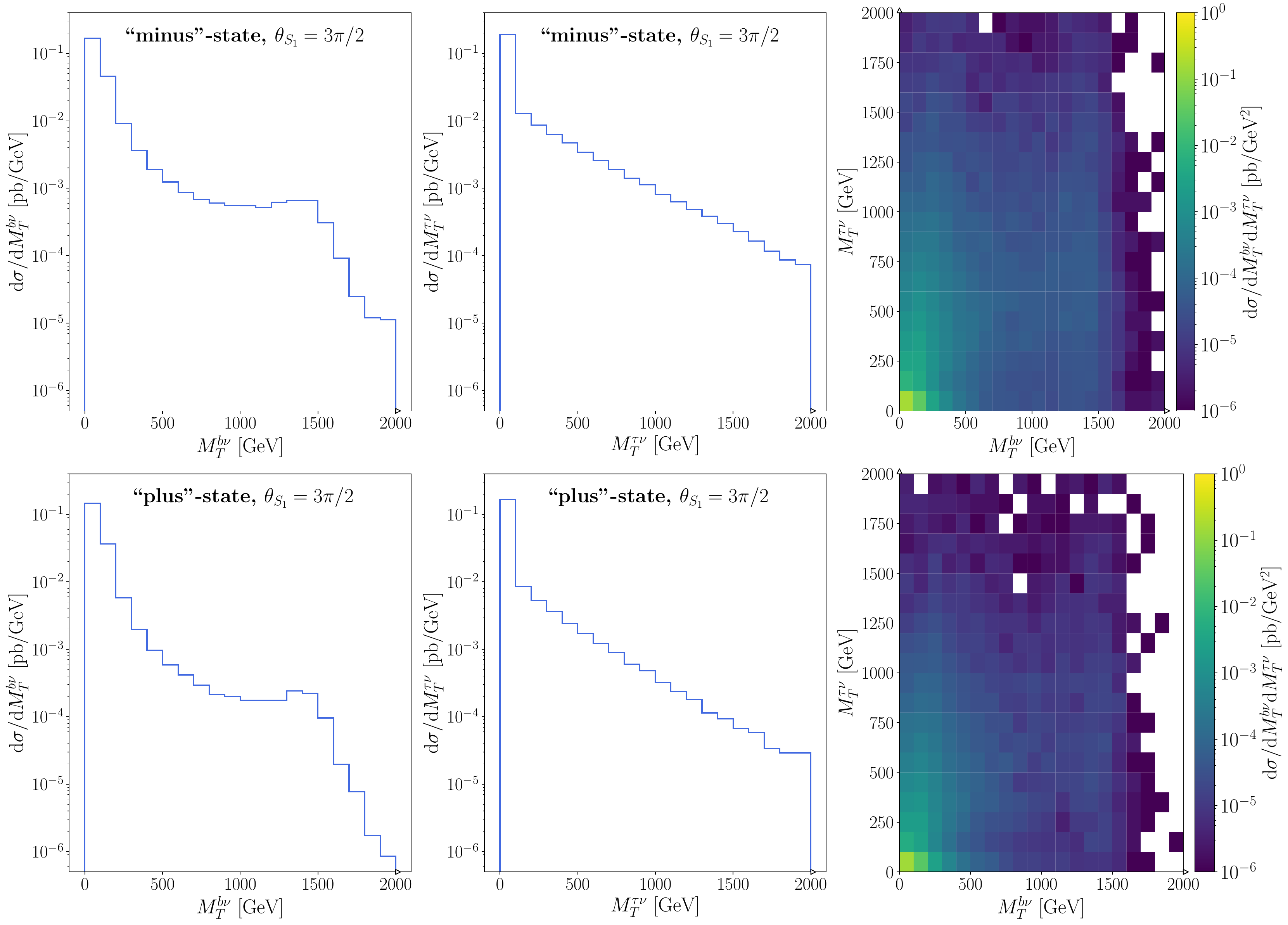}
    \caption{A comparison of signal events with a phase of $\theta_{S_1}=3\pi/2$ events from the ``minus''-state (top) and the ``plus'' state (bottom) for the transverse mass of the $b-\nu$ system on the left, $\tau-\nu$ system in the middle, and the 2D distribution of the transverse masses on the right.
    }
    \label{fig:mt_sig3pi2} 
\end{figure}

Here we study the differential cross-sections projected onto $M_T^{\tau\nu}$ or $M_T^{b\nu}$ plane, and distributed over the $M_T^{\tau\nu}$ and $M_T^{b\nu}$ plane. We reiterate that this study is done at parton level. The $W+b$ jet background sample is  illustrated in Fig.~\ref{fig:mt_bkg}, and a sample for $\theta_{S_1}=3\pi/2$ is illustrated in Fig.~\ref{fig:mt_sig3pi2} -- the phase assignment generating a maximum asymmetry in Fig.~\ref{fig:SigandBG}. In both figures, the top panel represents the ``minus''-state and the bottom panel represents the ``plus''-state, as defined above. The leftmost panels presents a variation of the cross-section with $M_T^{b\nu}$, while the middle is the cross-section with respect to $M_T^{\tau\nu}$. The rightmost panels show the two-dimensional transverse mass distributions, akin to a Dalitz plot for $1\to 3$ decays. 

In both plots, the signal structure clearly differs from the background, in particular, the signal leftmost panels show two noticeable peaks: one near the $W$ boson mass and another around the LQ mass. In comparison, the background only presents the first peak at the $W$ pole. Note that, when comparing the signal differential distributions for final state combinations, the new particle around $1.5$ TeV is observed to couple to the $b\nu$ final state rather than the $\tau\nu$, unlike a SM particle. 
On the other hand, both signal and background, demonstrate a small visible-by-eye difference when comparing between the two charge states.  

\begin{figure}[t!]
    \centering
    \includegraphics[width=0.325\textwidth]{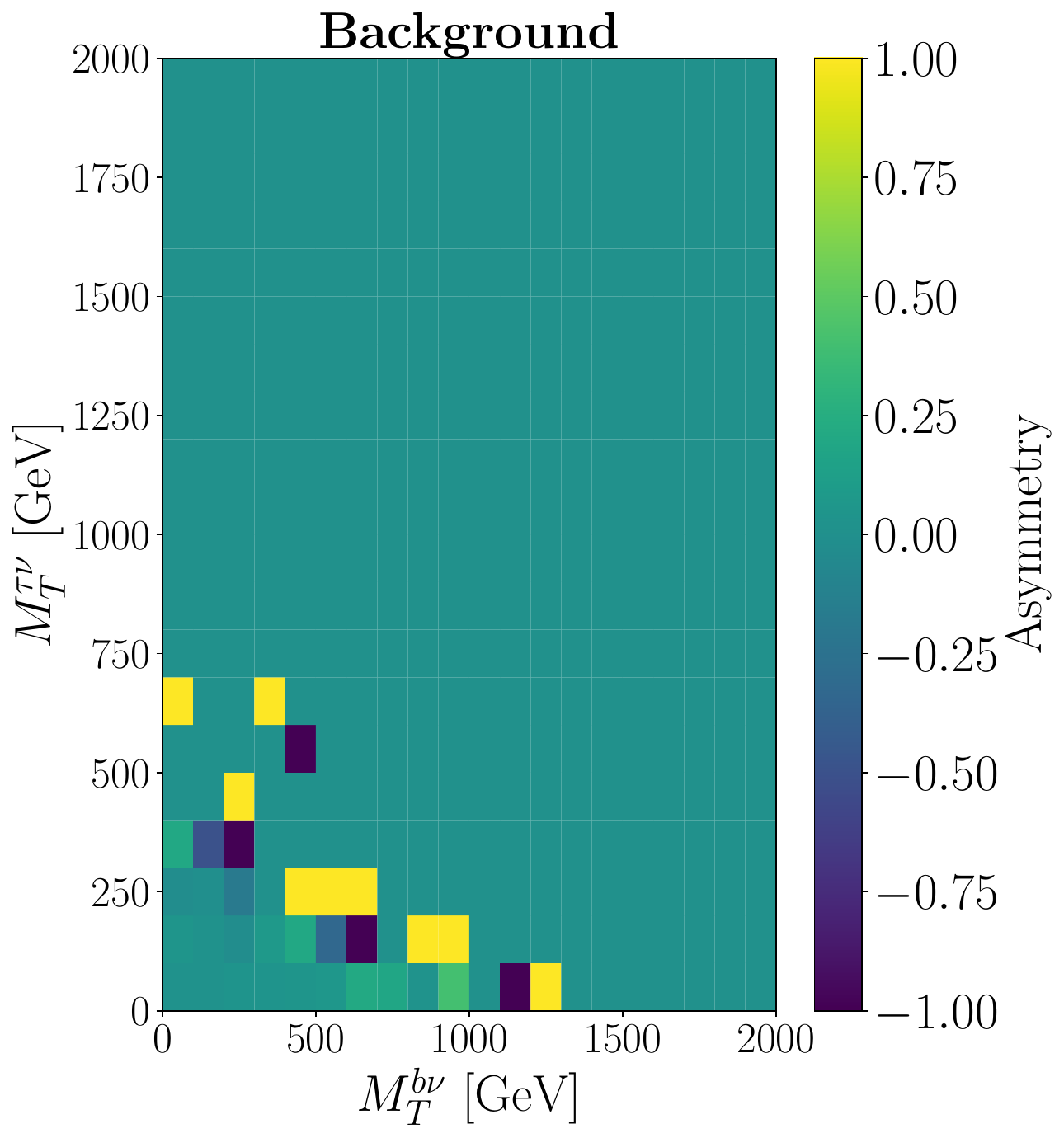}
    \hfill
    \includegraphics[width=0.325\textwidth]{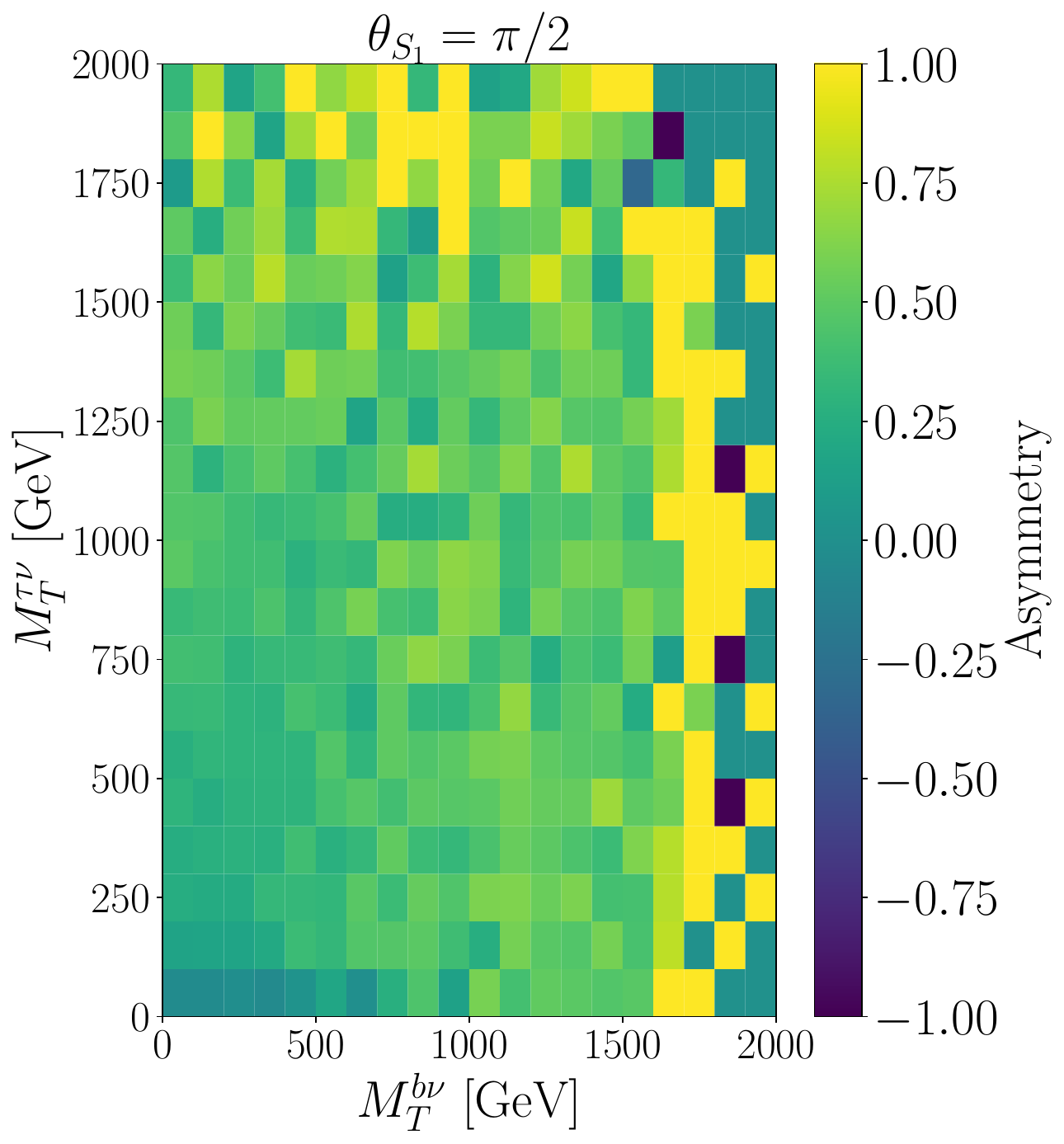}
    \hfill
    \includegraphics[width=0.325\textwidth]{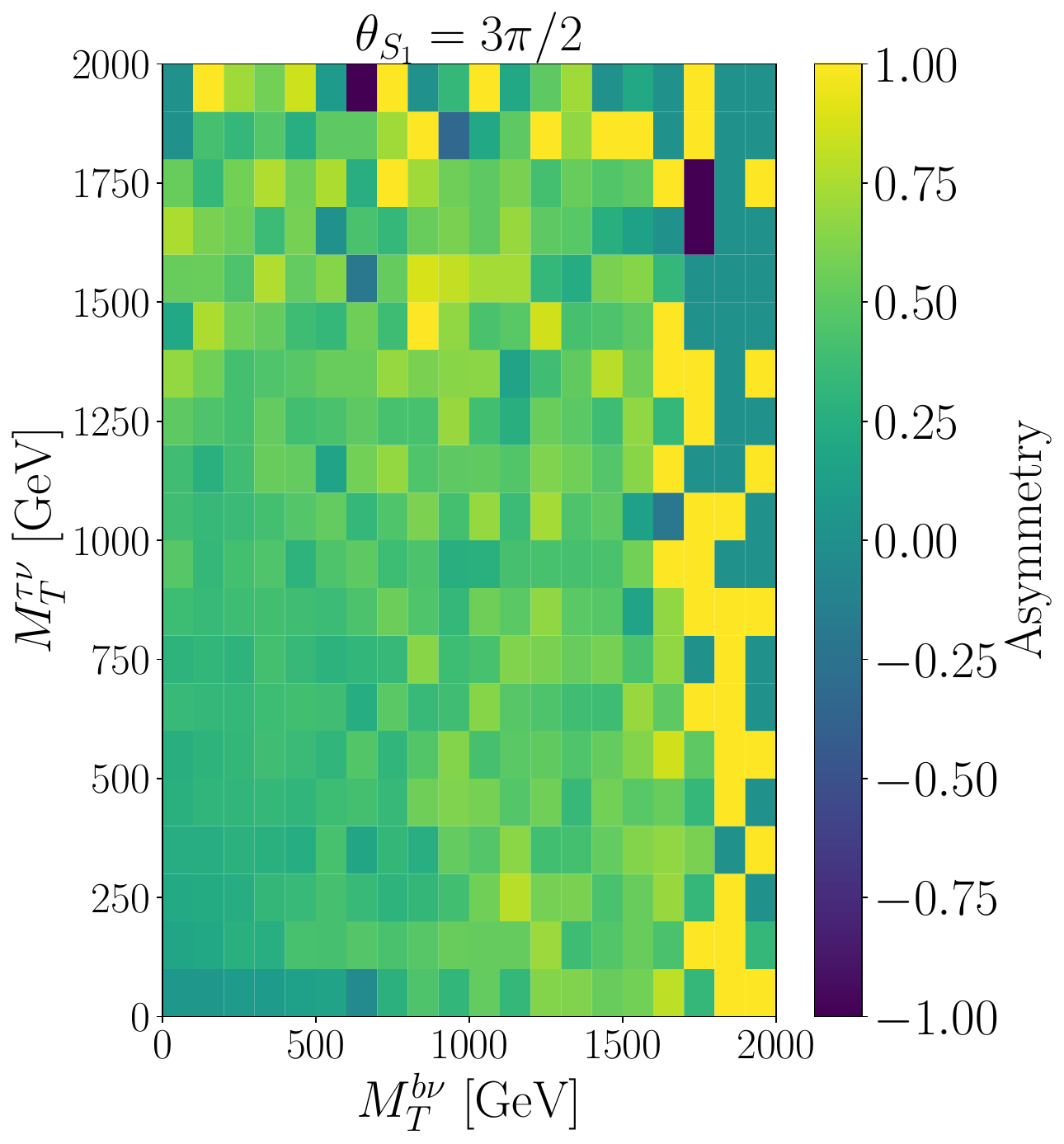}
    \caption{The asymmetry as a function of transverse masses between the two charge states for the background (left), signal with a phase of $\theta_{S_1}=\pi/2$ (middle), and signal with a phase of $\theta_{S_1}=3\pi/2$ (right). }
    \label{fig:mt_asym}
\end{figure}

In  Fig.~\ref{fig:mt_asym} we illustrate two-dimensional asymmetry distributions between the two charge states for the background (leftmost panel) and the signal for $\theta_{S_1}=\pi/2$ (middle panel) and $\theta_{S_1}=3\pi/2$ (rightmost panel), corresponding to the minimum and maximum asymmetry values from Sec.~\ref{ssec:valid}. Visually, the difference between these two signal phase assignments is subtle. From Fig~\ref{fig:SigandBG} we observe that the averaged asymmetry is shifted overall between these two phases, but the details revealing the influence of the two pole structures is challenging to distinguish between these two samples.  A more sophisticated analysis may enable the ability to enable a more careful spectroscopic study, particularly through the implementation of cuts around projected signal regions which may be utilized to enhance the asymmetry. 

\subsection{Particle-level analysis}
\label{sec:particle_level}

\begin{figure}[t!]
\begin{center}
\includegraphics[width=0.45\linewidth]{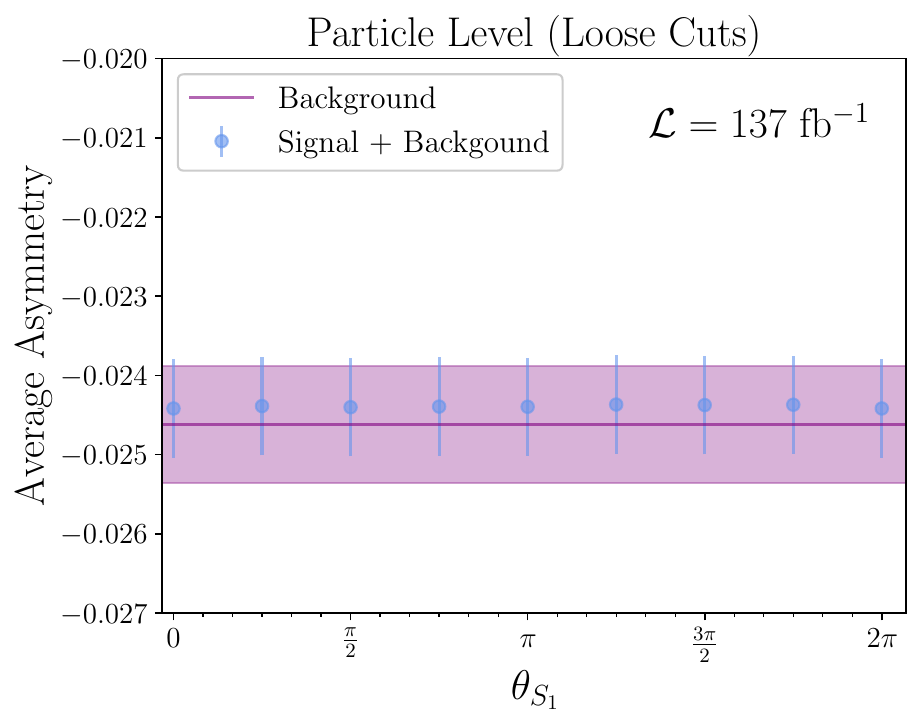}
\hfill
\includegraphics[width=0.45\linewidth]{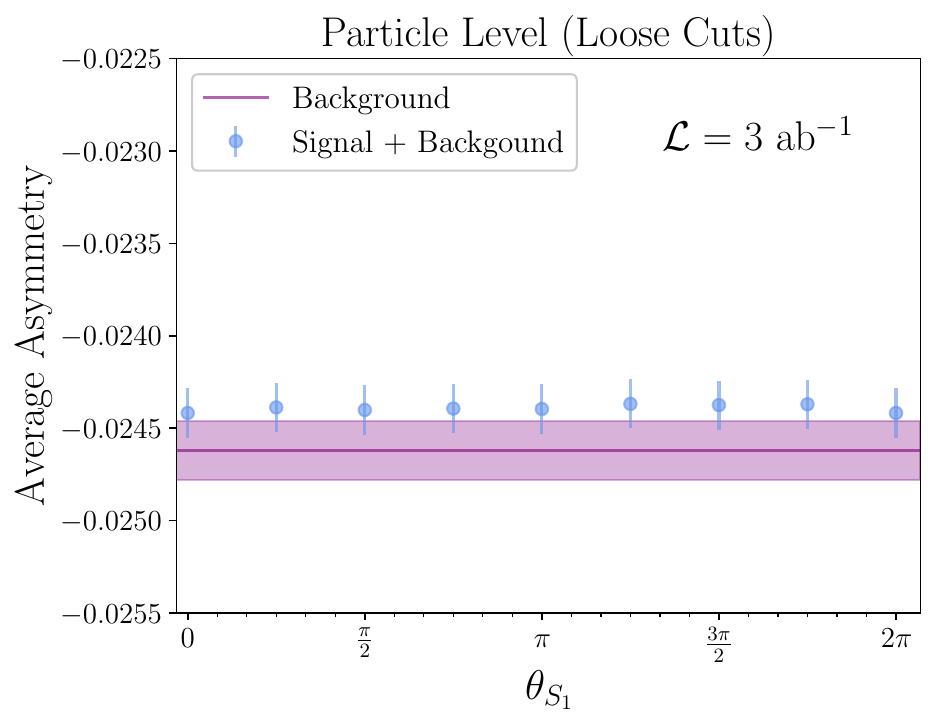}
\caption{Average asymmetry of the $pp \to b\tau\nu_\tau$ signal and $W$+jets background for different values of $\theta_{S_1}$. The signal is simulated, showered and hadronized for a for a luminosity of $137$ fb$^{-1}$ (left) and $3$ ab$^{-1}$ (right). We considered a LQ mass of $M_S = 1.5$ TeV,  $|C_{V_L}|=1$, and minimal transverse momentum cuts, listed in Tab.~\ref{tab:MinimalCuts}. The uncertainty bands represent the statistical uncertainties for a given luminosity. This scenario includes no top tagging for reduction of $t\bar{t}$ background.}
\label{fig:asym_showered}
\end{center}
\end{figure}

We extend the above by considering the effects of the parton shower~\cite{Schumann:2007mg}, hadronization~\cite{Chahal:2022rid}, and the decay of the tau~\cite{Siegert:2006xx,Laubrich:2006aa}.
For this analysis, we focus solely on the hadronic decays of tau leptons to mitigate the significant contribution to the background from $W$ decays to light leptons -- which are experimentally inseparable from the signal process.
These events are output into \texttt{HepMC3}~\cite{Buckley:2019xhk} files and analyzed using \texttt{Rivet}~\cite{Bierlich:2019rhm}. 
The $b$-tagger is run with a tagging efficiency of 70\%, a mistagging efficiency for $c$-jets of 10\%, and 1\% mistagging efficiency for light jets -- consistent with the running point for the ATLAS collaboration~\cite{2016}. 
The $\tau$ tagging efficiency is given as 60\% for one prong decays and 50\% for three prong decays, with a misidentification rate of 1/70 for one prong jets and 1/700 for three prong jets as given by the tight cuts from Refs.~\cite{ATLAS:2024tzc, ATL-PHYS-PUB-2019-033}.
The jets, taus, and missing transverse momentum are smeared according to the ATLAS run 2 benchmark point to mimic the impact of detector effects using the \texttt{Rivet} implementation~\cite{Buckley:2019stt}. The jets and the $\tau$ are smeared using a Gaussian distribution using a parameterization based on Ref.~\cite{ATLAS:2019oxp}.
The parameterization of the missing transverse momentum is smeared using a Gaussian distribution based on a resolution given in Ref.~\cite{ATLAS:2018txj}.
Jets are defined using the anti-kt algorithm using \texttt{FastJet}~\cite{Cacciari_2012, Cacciari_2006} with $\Delta R=0.4$ and $p_T>$ 25 GeV.
Events are required to have at least one $\tau$-jet, exactly one $b$-jet, and ${\rm MET} > 25$ GeV.
The charge state is determined by measuring the charge of the tau-jets,
and assume that the charge of the $b$-jet is the opposite of the tau-jet.
The charge identification of hadronic taus is 99\% for channels with a single charged hadron and 70\% for channels with three charged hadrons~\cite{CMS:2022prd}.
Taking a weighted average of the branching ratios, this gives an on average charge identification rate of roughly 92.5\%.
Since this is only a proof-of-principle study, and the implementation of the detector effects is only approximate, it is appropriate to take the simplifying assumption that the charge identification is 100\%. This should not have any drastic impact on the conclusions of this work.
Limiting the number of $b$-jets to exactly one is done to drastically reduce the background from $t\bar{t}$ events.

As before, we investigate looking at the asymmetry between the two charge states after parton showers, hadronization, and detector smearing. The average asymmetry for the signal and background is illustrated in Fig.~\ref{fig:asym_showered}. Again, the uncertainty bands only represent the statistical uncertainty. We represent results for the amount of data currently collected at the LHC (137 fb${}^{-1}$) on the left, and the final luminosity expected at the end of the HL-LHC (3 ab${}^{-1}$) on the right. We find that the central value for the asymmetry for the background is dominated by $W$+light jets, causing a significant shift to the expected asymmetry from Fig.~\ref{fig:SigandBG}. 

In Fig.~\ref{fig:asym_showered}, we show the default scenario with very limited cuts. In this situation, we see that the benchmark point would only reach a significance of about 1$\sigma$ for the presence of an asymmetry at the end of the HL-LHC, although the exact phase information is washed out. 

\subsubsection{Machine Learning Implementation}
\begin{figure}[t!]
\begin{center}
\includegraphics[width=0.5\linewidth]{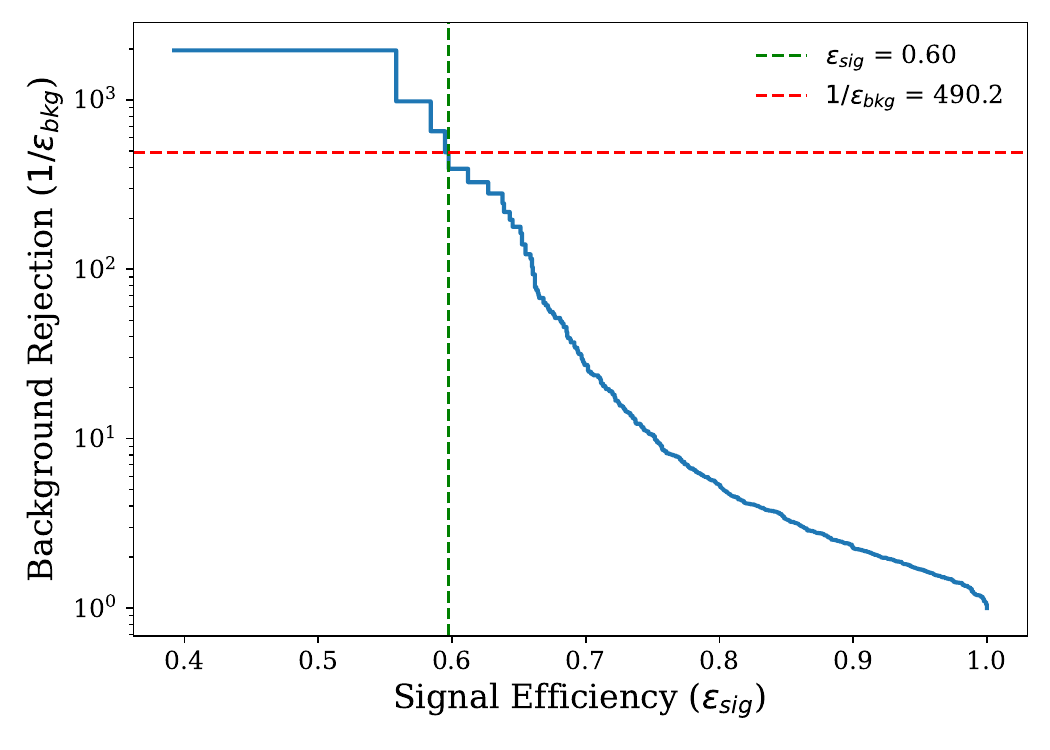}
\caption{The background rejection as a function of the signal efficiency is given in the blue curve. The green vertical dashed line indicates the working point chosen for this work, and corresponds to a signal efficiency of \(\varepsilon_{\text{sig}} = 0.60\), while the red horizontal dashed line marks the corresponding background rejection of \(1/\varepsilon_{\text{bkg}} \approx 490.2\).
    }
\label{fig:classifier}
\end{center}
\end{figure}

To address the limitations inherent in the simple cut-based approach outlined above, we propose the use of a machine learning (ML) framework to enhance the discrimination between signal and background. Neural networks, in particular, have emerged as powerful tools in this domain, with demonstrated success in improving the sensitivity of searches for BSM physics~\cite{Karagiorgi:2021ngt, ATLAS:2025rbr, Huang:2025ziq, ATLAS:2024rpl, Leigh:2024chm, Plehn:2022ftl, Kasieczka:2021xcg}
As a first step, we employ a simple fully connected neural network to establish a transparent baseline, thereby illustrating the potential of ML techniques even in the absence of complex architectural designs. Within this framework, we train a binary classifier to distinguish between signal and background events. The background samples considered include $W+$jets, $Z+$jets and $t\bar{t}+$jets processes. We do not include the $W+b$-jet SM process, as the asymmetry observable alone is sufficient to isolate contributions from new physics to this process, as shown in Sec.~\ref{sec:partonlevel}. Furthermore, the background is dominated by the processes with light-jets that are misidentified as $b$-jets. We employed the \texttt{Scikit-Learn} library~\cite{pedregosa2018scikitlearnmachinelearningpython} to implement a Multi-Layer Perceptron binary classifier. The network architecture consists of 10 input neurons: corresponding to the four-momentum of the $\tau$-jet and the $b-$jet ($4+4$ neurons), as well as the x and y components of the missing transverse momentum. The model consists of two hidden layers with 64 and 32 neurons, respectively, and is trained for up to 1000 epochs, with early stopping~\cite{bai2021understandingimprovingearlystopping} employed to prevent overfitting. Non-linearity is introduced via the ReLU~\cite{10.5555/3104322.3104425} activation function in the hidden layers, while the output layer uses a sigmoid~\cite{nonlinref} activation, suitable for binary classification. The training performance is evaluated using the binary cross-entropy loss function. After training, we assess the ability of the classifier to distinguish between signal and background by computing the signal efficiency and background rejection, defined as
\begin{align}
\varepsilon_{\text{sig}} &= \frac{N_{\text{sig}}^{\text{selected}}}{N_{\text{sig}}^{\text{total}}}, \hspace{1cm}
\frac{1}{\varepsilon_{\text{bkg}}} = \frac{N_{\text{bkg}}^{\text{total}}}{N_{\text{bkg}}^{\text{selected}}},
\end{align}
where $N_{\text{sig}}^{\text{selected}}$ ($N_{\text{bkg}}^{\text{selected}}$) denotes the number of signal (background) events classified as signal by the model, and $N_{\text{sig}}^{\text{total}}$($N_{\text{bkg}}^{\text{total}}$) represents the total number of signal (background) events, regardless of the classifier’s decision. Figure~\ref{fig:classifier} illustrates the trade-off between signal efficiency ($\varepsilon_{\text{sig}}$) and background rejection($1/\varepsilon_{\text{bkg}}$) achieved by the trained classifier. The blue solid curve shows the classifier's performance across different decision thresholds.

\begin{figure}[t!]
\begin{center}
\includegraphics[width=0.45\linewidth]{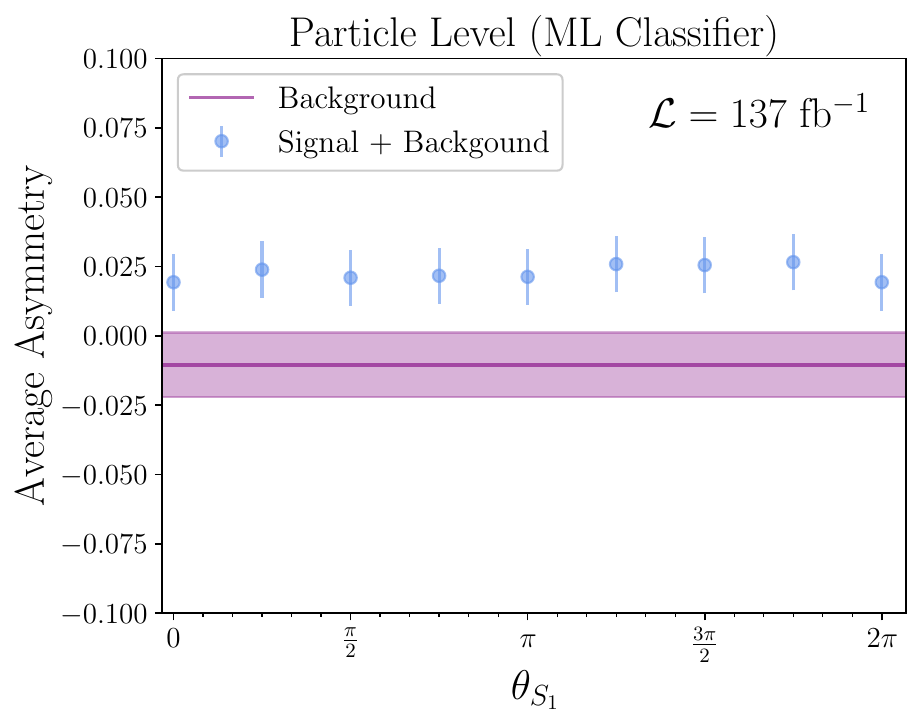}
\hfill
\includegraphics[width=0.45\linewidth]{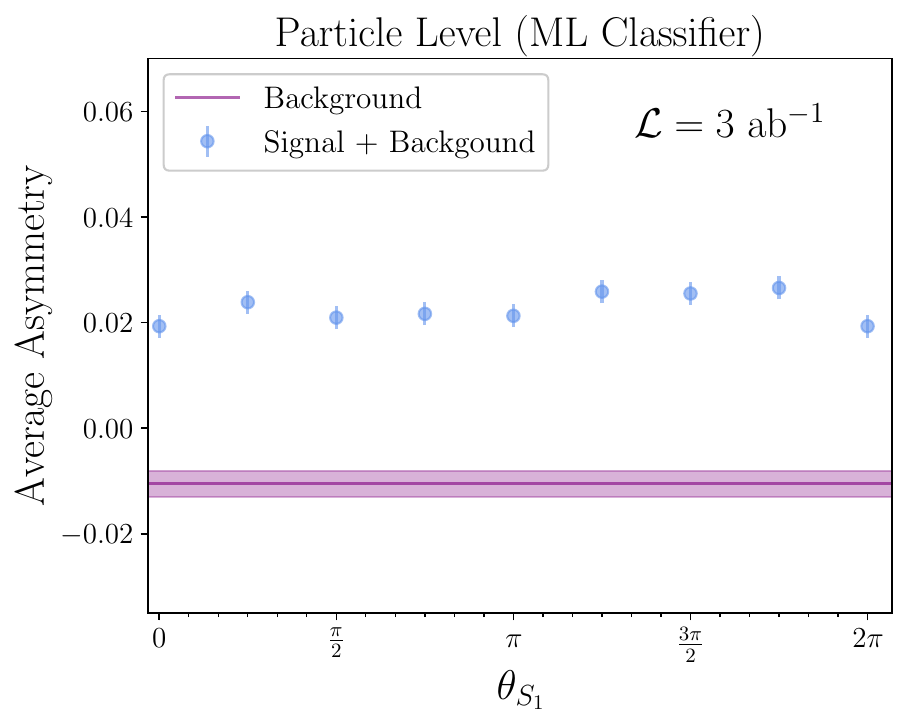}
\caption{Average asymmetry of the $pp \to b\tau\nu_\tau$ signal and $W$+jets and $Z$+jets backgrounds for different values of $\theta_{S_1}$. The signal is simulated, showered and hadronized for a for a luminosity of $137$ fb$^{-1}$ (left) and $3$ ab$^{-1}$ (right). We considered a LQ mass of $M_S = 1.5$ TeV,  $|C_{V_L}|=1$, and minimal transverse momentum cuts, listed in Tab.~\ref{tab:MinimalCuts}.
The uncertainty bands represent the statistical uncertainties for a given luminosity.
These plots include a reduction of the reducible backgrounds by the implementation of a binary classifier misidentifying 1 out of every 490
background events as a signal event.}
\label{fig:asym_ML}
\end{center}
\end{figure}

To improve our sensitivity to the asymmetry, we leverage our classifier choosing a working point of a signal efficiency of 60\%, which corresponds to a background rejection rate of 490 (\textit{i.e.} for every 490 background events, we accept one of them as if it was a signal event).
After applying this classifier to our data, we obtain a significant improvement to our sensitivity to the new physics in the asymmetry observable as shown in Fig.~\ref{fig:asym_ML}.
To estimate the overall significance of this result, we include a conservative 10\% systematic uncertainty, with which the significance of discriminating signal from background in this model is 2$\sigma$ for the current LHC dataset, and is improved to 8$\sigma$ for the HL-LHC.
Our results show that the LHC, using the currently available dataset, has the potential to exclude the presence of such an asymmetry, thereby tightening existing constraints on these BSM scenarios.

Unfortunately sensitivity to the exact value of the CP-violating phase inherent to the model remains limited, even with the projected dataset from the High-Luminosity LHC. 
This is primarily due to the smearing of the transverse mass caused by the secondary neutrino from tau decay. 
Although a full extraction of model parameters is beyond the scope of our analysis, we briefly highlight potential avenues for further studies of this experimental asymmetry. For the process studied in this work, the use of more sophisticated techniques to handle final states with multiple neutrinos, such as the $M_{T2}$ observable~\cite{Lester:1999tx,Barr:2003rg}, may help recover lost information. Moreover, applying an analogous procedure and asymmetry measure to a different $2\to3$ process without such a difficult invisible final state could preserve the oscillatory behavior with the weak-CP phase and enhance sensitivity to model parameters.  As such, we hope to motivate future investigations to fully exploit the potential of this asymmetry observable for parameter determination and limit setting.

\section{Conclusion}
\label{sec:clonclusion}

Driven by the need for a deeper understanding of new sources of CP violation and their expression in BSM frameworks, we propose a novel observable based on the kinematic asymmetry of the $2 \to 3$ scattering process. Our study focuses on the interference between SM charged-current decays and BSM contributions in high-energy colliders, particularly manifest in a charge asymmetry of the $pp \to b\tau\nu_\tau$ process. To achieve this, we utilize a model that extends the SM by introducing a scalar leptoquark, $S_1$. 

In order to investigate the impact of the asymmetry as a relevant parameter at collider studies, we perform a numerical analysis at both parton and particle level for the $pp \to b\tau\nu_\tau$ signal and the corresponding backgrounds. 
The parton-level analysis begins with an initial investigation of the asymmetry scanning over several values of the $C_{V_L}$ phase ($\theta_{S_1}$). It is observed that the signal is distinctly separable from the background and exhibits the expected oscillatory behavior showing a maximum and minimum average asymmetry values of $ \theta_{S_1}  = 3\pi/2$ and $ \theta_{S_1} = \pi/2$, respectively, demonstrating that pole effects indeed dominate the asymmetry behavior. Furthermore, we performed a comparison of signal and background events from the `plus' and  `minus' charge state (as defined by the charge of the final-state tau) for the transverse mass of the $b-\nu$ and $\tau-\nu$ system, including the two-dimensional distribution of the transverse masses. Here the signal demonstrates two peaks at the $W$ and LQ masses (unlike the background, which only presents the $W$ peak) revealing the presence of a $1.5$ TeV particle coupling to $b\nu$. A closer examination of the two-dimensional transverse mass distribution reveals a clear separation between the signal benchmarks and the background. However, distinguishing between the two signal phase assignments remains challenging, highlighting the need for more advanced techniques to achieve this separation. 

At particle level, the analysis is dominated by the $W$+jets background -- primarily due to the use of very loose selection cuts. Moreover, the oscillatory behavior of the CP-sensitive asymmetry is diluted by parton showering, hadronization, detector resolution effects, and by the presence of a secondary neutrino from tau decays. To circumvent these limitations, we implement a binary classifier to discriminate between signal and background events, with the latter including $W$+jets, $Z$+jets, and $t\bar{t}$+jets. We fix the signal efficiency at 60\%, which corresponds to accepting one background event out of every 490 rejected. 
Repeating the analysis with this classification step reveals a significant enhancement in sensitivity to the presence of new physics through the asymmetry observable. Despite the challenges posed by detector effects and secondary neutrinos, our results show that even with minimal selection cuts, the presence of BSM-SM interference can be resolved with the aid of Machine Learning techniques, reaching a significance of 2$\sigma$ with the current dataset and up to 8$\sigma$ for the HL-LHC. 
Ultimately, we demonstrate the promise of such interference-sensitive observables in three-body final states and lay the groundwork for future efforts to improve parameter extraction in BSM scenarios.

\acknowledgments
\noindent We would like to thank Rodolfo Capdevilla his contributions in the initial development of this work and in establishing this collaboration. We also thank Jonathan Kriewald useful discussion, and for his assistance with FeynRules. We would like to thank Daniel Hayden for useful comments pertaining to $\tau$-jet tagging. IB would like to thank David Curtin, Yuval Grossman, Roni Harnik for useful discussion. This manuscript has been authored in part by  Fermi Forward Discovery Group, LLC under Contract No. 89243024CSC000002 with the U.S. Department of Energy, Office of Science, Office of High Energy Physics. The work of J.I. was supported by the U.S. Department of Energy, Office
of Science, Office of Advanced Scientific Computing Research, Scientific Discovery through Advanced Computing
(SciDAC-5) program, grant “NeuCol”. The work of I.B. was performed in part at the Aspen Center for Physics, which is supported by a grant from the Alfred P. Sloan Foundation (G-2024-22395). The work of K.T is supported by DOE Grant KA2401045.
\appendix

\section{\texorpdfstring{Interference effects in direct observables and $\mathcal{A}_\text{CP}$}{Int Acp}}
\label{App:interf}
Here we explicitly detail the generation of an asymmetry in interference between SM and BSM physics, following a similar prescription to Ref.~\cite{Nowakowski:1989ju}. Recall the following Eq.~\eqref{eq:BSMSM} which highlights the interference generated between SM and BSM contributions to a process. Parameterizing the contributions from the SM and the BSM amplitudes as the following
\begin{align}
\mathcal{M}_\text{BSM} = |\mathcal{M}_\text{BSM}|e^{i\delta_\text{BSM}}e^{i\theta_\text{BSM}},\; \mathcal{M}_\text{SM} = |\mathcal{M}_\text{SM}|e^{i\delta_\text{SM}}e^{i\theta_\text{SM}}. 
\end{align}
where $\delta$ represents a CP-even phase, and $\theta$ represents a CP-odd phase from either contribution, and $\Delta \theta =\theta_\text{SM}-\theta_\text{BSM}, \Delta \delta =\delta_\text{SM}-\delta_\text{BSM}$
\begin{align}
|\mathcal{M}|^2= |\mathcal{M}_\text{BSM}|^2 &+|\mathcal{M}_\text{SM}|^2 + 2 |\mathcal{M}_\text{BSM}||\mathcal{M}_\text{SM}|\cos(\Delta\delta+\Delta\theta).\nonumber 
\end{align}
Now, consider the CP-conjugate process, 
\begin{align}
    |\overline{\mathcal{M}}|^2 = |\mathcal{M}_\text{BSM}|^2 &+|\mathcal{M}_\text{SM}|^2 + 2 |\mathcal{M}_\text{BSM}||\mathcal{A}_\text{SM}|\cos(\Delta\delta-\Delta\theta)\nonumber 
\end{align}
Therefore, the difference is given by 
\begin{align}
   &|\mathcal{M}|^2-  |\overline{\mathcal{M}}|^2 
   = -4 |\mathcal{M}_\text{BSM}||\mathcal{M}_\text{SM}|\sin(\Delta \delta) \sin (\Delta \theta)
\end{align}
and the asymmetry
\begin{align} \label{eq.ACP_express}   &\mathcal{A}_\text{CP} =  \frac{-2 |\mathcal{A}_\text{BSM}||\mathcal{A}_\text{SM}|\sin(\Delta \delta) \sin (\Delta \theta)}{|\mathcal{A}_\text{BSM}|^2 +|\mathcal{A}_\text{SM}|^2 +2|\mathcal{A}_\text{BSM}||\mathcal{A}_\text{SM}|\cos(\Delta \delta) \cos (\Delta \theta)}.
\end{align}
If the the CP-even phase difference is $\Delta \delta=\pm \pi/2$ (corresponding, perhaps, to a maximal difference in virtuality between the BSM and SM propagators) then 
\begin{align}
\mathcal{A}_\text{CP} \approx   \mp\frac{2 |\mathcal{A}_\text{BSM}||\mathcal{A}_\text{SM}|\sin (\Delta \theta)}{|\mathcal{A}_\text{BSM}|^2 +|\mathcal{A}_\text{SM}|^2 }. 
\end{align}
Moving away from this regime, the dependence on the CP-even phase difference re-enters and it will become important where (kinematically) the asymmetry is being sampled. This motivates considering an asymmetry in both total decay rates \emph{and} in differential decay rates. The focus of this differential decay rate interference in reference~\cite{Berger:2011wh} was tree-level $1\to 3$ decays, however as can be seen above, this asymmetry is not limited to decays but should also be relevant for tree-level scattering.

\section{EFT parameterization}
\label{appA:EFT}
    The contributions to $d_j\to u_i \ell \nu_\ell$ are given by the following effective interactions in the Weak Effective Theory (WET)~\cite{Aebischer:2017ugx}, where 
\begin{align}
    \mathcal{L}_\text{WET} \supset -\frac{4G_F}{\sqrt{2}}V_{ij}\sum_k C_k \mathcal{O}^k \label{eq:EFT}
\end{align}
with the sum indicating a sum over the operator basis defined below~\footnote{We have suppressed the Lorentz structure labels in Eq.~\eqref{eq:EFT} above for brevity. }
\begin{eqnarray}
\label{eq:EFTbasis}
\begin{aligned}
&(\mathcal{O}_{S_L})_{ji\ell} =(\overline{u_i} P_L d_j)(\overline{\ell}P_L \nu_\ell),\\
&(\mathcal{O}_{T})_{ji\ell}= (\overline{u_i} \sigma^{\mu\nu} P_L d_j)(\overline{\ell}\sigma_{\mu\nu}P_L\nu_\ell),\\
&(\mathcal{O}_{S_R})_{ji\ell}= (\overline{u_i} P_R d_j)(\overline{\ell}P_L \nu_\ell),\\
&(\mathcal{O}_{V_L})_{ji\ell}= (\overline{u_i}\gamma^\mu P_L d_j)(\overline{\ell}\gamma_\mu P_L\nu_\ell).
\end{aligned}
\end{eqnarray}
and the associated $C_{ji\ell}$ are the new-physics Wilson coefficients (WCs). The normalization here defined so that in the SM the only nonzero WC for this process is $C_{V_L}^\text{SM}=1$. Here $\sigma_{\mu\nu} = \frac{i}{2} [\gamma_\mu, \gamma_\nu]$. Note that the LQ interactions need not generally conserve lepton flavour, although the $\nu_\ell -\ell$ interactions maximize interference with the SM processes, so for this work we will solely consider lepton-flavour conserving contributions. 

At the high-energy scales probed by $pp$ collisions, it is appropriate to consider the parton-level process $d_j\to u_i \ell \nu_\ell$ as occurring via a standard-model effective theory~(SMEFT) interaction, parameterized by
\begin{align}
    \mathcal{L}_\text{SMEFT} \supset \frac{1}{v^2 }\sum_i C_k \mathcal{O}^k \label{eq:SMEFT}
\end{align}
where $v\approx 246$ GeV is the SM Higgs vacuum expectation value, and the WCs scale as $C_k\sim v^2/\Lambda^2$ and $\Lambda$ is the new-physics scale. We utilize the Warsaw basis for this EFT, with the relevant operators given by
\begin{align}
    &(\mathcal{O}^{(3)}_{lq})_{\ell ij}= (\overline{L_L} \gamma_\mu \tau^I L_L)(\overline{Q_L^i} \gamma^\mu \tau^I Q_L^j)\\
    &(\mathcal{O}_{ledq})_{\ell ij}= (\overline{L_L} e_R)(\overline{d_R^i} Q_L^j)\\
    &(\mathcal{O}^{(1)}_{lequ})_{\ell ij}= (\overline{L_L^a} e_R)\epsilon_{ab} (\overline{Q^{i\;b}_L} u_R^j)\\
    &(\mathcal{O}^{(3)}_{lequ})_{\ell ij}= (\overline{L_L^a} \sigma_{\mu\nu} e_R)\epsilon_{ab} (\overline{Q^{b\; i}_L}  \sigma^{\mu\nu}u^j_R)
\end{align}
where $\tau^I$ are the Pauli matrices and $\epsilon_{ab}$ is the Levi-Civita tensor acting over $SU(2)_L$ space. Here we have suppressed flavour indices for simplification. Here $L_L$ and $Q_L$ are the $SU(2)_L$ SM lepton and quark doublets, respectively. We have suppressed flavour indices at this stage, but will reintroduce them below when considering the matching between coefficients of these two bases.

To compare and contrast the effects of models at the high-energy (SMEFT) scale and the low-energy (WET, $\mu_B$) scale we need to consider the running and matching of the WCs between these two bases. The matching is defined at $\mu=m_W$ to be
\begin{align}
&C_{S_L}^{ji\ell}= -\frac{V_{kj}}{2 V_{ij}}[C^{(1)}_{lequ}]^*_{\ell ki},\;\;
C_{T}^{ji\ell}= -\frac{V_{kj}}{2 V_{ij}}[C^{(3)}_{lequ}]^*_{\ell k i},\\
&C_{S_R}^{ji\ell}= -\frac{1}{2 V_{ij}}[C_{ledq}]^*_{\ell ji},\;\;\;\;C_{V_L}^{ji\ell}= -\frac{V_{kj}}{V_{ij}}[C^{(3)}_{lq}]_{\ell ik}
\end{align}
where there is a summation only over the index $k$.~\footnote{The operator $C_{V_L}$ also should have a further correction from any modification to the $W$-quark coupling, which doesn't obtain corrections from the LQ models we consider and so we neglect this for the purpose of our discussion. Extracting the CKM parameters in the EFT is a an important and nontrivial exercise, but for our purposes we will assume the results from standard CKM fits~\cite{Descotes-Genon:2018foz}.}

\section{\texorpdfstring{The Scalar Leptoquark, $S_1$}{Scalar LQ S1}}

The interactions of this LQ with SM fermions are given by the following\footnote{Here, $L$ is the SM lepton isospin doublet, $Q$ is the SM quark doublet, $u$ and $e$ are the SM up-quark and charged-lepton isosinglets, respectively.}
\begin{equation}
 \mathcal{L}_{S_1} \supset \lambda_L^{ij} \overline{L_i^c} P_L Q_j S_1^* + \lambda_R^{ij} \overline{e^c_i} P_R u_j S_1^*+ \text{h.c.},
\end{equation}
which, after EWSB, can be rewritten as
\begin{equation}
\begin{aligned}
\label{eq:S1int}
\mathcal{L}_{S_1} \supset  x_{S_1}^{ij} \overline{e^c_i} P_L u_j S_1^* &+ z_{S_1}^{ij} \overline{\nu^c_i} P_L d_j S_1^* + y_{S_1}^{ij} \overline{e^c_i} P_R u_j S_1^* + \text{h.c.}
\end{aligned}
\end{equation}
where we have defined \footnote{We chose  the order of the indices in $\lambda^{ij}$ such that $i$ refers to lepton flavour, then $j$ to quark flavour, so that Lepto-Quark can act
as a mnemonic.}
\begin{eqnarray}
   & z_{S_1}^{ij} = [\mathfrak{L}_e^\text{T}\lambda_L \mathfrak{L}_d]_{ij},\;\; x_{S_1}^{ij}=-[\mathfrak{L}_e^\text{T}\lambda_L \mathfrak{L}_u]_{ij},\;\;y_{S_1}^{ij}=[\mathfrak{R}_e^\text{T}\lambda_R \mathfrak{R}_u]_{ij}.
\end{eqnarray}
Here $\mathfrak{L}_a, \mathfrak{R}_b$ are the rotation matrices transforming between the electroweak gauge and mass eigenstates of the SM fields. Therefore, the couplings $z, x$ to left-handed quark fields are not uniquely defined, and are related by the CKM matrix via $z^{ij}=-x^{ik} V_{kj}$.  If we adopt the down-aligned basis (i.e. fix the mixing matrix of the left-handed down-type quarks to be diagonal) then we have 
\begin{align} &\mathfrak{R}_e =\mathfrak{R}_u =\mathfrak{R}_d=\mathbf{1},  \mathfrak{L}_e = \mathfrak{L}_d =\mathbf{1},\text{so that}\;\;\;V_{\rm CKM} =\mathfrak{L}_u^\dagger.\end{align}
Therefore, following from above, the couplings are  
\begin{eqnarray}
    x_{S_1}^{ij}=-\lambda_L^{ik} V^\dagger_{kj},&\quad
y_{S_1}^{ij}=\lambda_R^{ij},  \quad
    z_{S_1}^{ij} = \lambda_L^{ij},
\end{eqnarray}
so that 
\begin{equation}
\begin{aligned}
\mathcal{L}_{S_1} 
\supset \lambda_L^{ij} &\left[\overline{\nu^c_i} P_L d_j  - V^\dagger_{jk} \overline{e^c_i} P_L u_k \right]S_1^*  
+ \lambda_R^{ij} \overline{e^c_i} P_R u_j S_1^* + \text{h.c.} \nonumber 
\end{aligned}
\end{equation}
To address the anomalies in $b\to c \tau\nu$, a minimal coupling setup consists of fixing $(\lambda_{L}^{33},\lambda_{L}^{32}, \lambda_{R}^{32})$ to nonzero values, therefore generating the following effective interactions (corresponding to dimension-6 operators in the WET EFT basis, as defined in Eq.~\eqref{eq:EFTbasis} of App.~\ref{appA:EFT}),
    \begin{align}
C^{S_1}_{S_L} (M_{S_1}) &= -4 C_{T}(M_{S_1}) = \frac{(\lambda_R^{32})^* \lambda_L^{33}}{4\sqrt{2}G_FV_{cb} M_{S_1}^2} \label{CVLCSL} ,\\
C_{V_L}^{S_1}(M_{S_1}) 
&= \frac{\lambda_L^{33}\left[(\lambda_L^{33})^* V_{cb}+(\lambda_L^{32})^*V_{cs}\right]}{4\sqrt{2}G_FV_{cb} M_{S_1}^2}. \nonumber
\label{eq:CVL_def_2}
\end{align}
\newpage 
\bibliography{biblio}

\end{document}